\documentclass[prb,twocolumn,showpacs,aps,superscriptaddress,floatfix]{revtex4}
\usepackage{amsmath}
\usepackage{amssymb}
\usepackage{bm}
\usepackage{graphicx}
\usepackage{color}
\usepackage{hyperref}
\usepackage{verbatim}

\begin{document}

\title{Spin-valley half-metal in systems with Fermi surface nesting}

\author{A.L. Rakhmanov}
\affiliation{Theoretical Quantum Physics Laboratory, RIKEN Cluster for
Pioneering Research, Wako-shi, Saitama 351-0198, Japan }
\affiliation{Institute for Theoretical and Applied Electrodynamics, Russian Academy of Sciences, Moscow, 125412 Russia}
\affiliation{Moscow Institute for Physics and Technology (State
University), Dolgoprudnyi, 141700 Russia}
\affiliation{Dukhov Research Institute of Automatics, Moscow, 127055 Russia}

\author{A.O. Sboychakov}
\affiliation{Theoretical Quantum Physics Laboratory, RIKEN Cluster for
Pioneering Research, Wako-shi, Saitama 351-0198, Japan }
\affiliation{Institute for Theoretical and Applied Electrodynamics, Russian Academy of Sciences, Moscow, 125412 Russia}

\author{K.I. Kugel}
\affiliation{Institute for Theoretical and Applied Electrodynamics, Russian Academy of Sciences, Moscow, 125412 Russia}
\affiliation{National Research University Higher School of Economics, Moscow, 101000 Russia}

\author{A.V. Rozhkov}
\affiliation{Theoretical Quantum Physics Laboratory, RIKEN Cluster for
Pioneering Research, Wako-shi, Saitama 351-0198, Japan }
\affiliation{Institute for Theoretical and Applied Electrodynamics, Russian
Academy of Sciences, Moscow, 125412 Russia}
\affiliation{Moscow Institute for Physics and Technology (State
University), Dolgoprudnyi, 141700 Russia}
\affiliation{Skolkovo Institute of Science and Technology, Skolkovo
Innovation Center, Moscow, 143026 Russia}

\author{Franco Nori}
\affiliation{Theoretical Quantum Physics Laboratory, RIKEN Cluster for
Pioneering Research, Wako-shi, Saitama 351-0198, Japan }
\affiliation{Department of Physics, University of Michigan, Ann Arbor, MI
48109-1040, USA}

\begin{abstract}
Half-metals have fully spin polarized charge carriers at the Fermi surface.
Such polarization usually occurs due to strong electron--electron
correlations. Recently [Phys. Rev. Lett. {\bf{119}}, 107601 (2017)], we
have demonstrated theoretically that adding (or removing) electrons to
systems with Fermi surface nesting also stabilizes the half-metallic states
even in the weak-coupling regime. In the absence of doping, the ground
state of the system is a spin or charge density wave, formed by four nested
bands. Each of these bands is characterized by charge (electron/hole) and
spin (up/down) labels. Only two of these bands accumulate charge carriers
introduced by doping, forming a half-metallic two-valley Fermi surface.
Analysis demonstrates that two types of such half-metallicity can be
stabilized. The first type corresponds to the full spin polarization of the
electrons and holes at the Fermi surface. The second type, with
antiparallel spins in electron-like and hole-like valleys, is referred to
as a ``spin-valley half-metal'' and corresponds to the complete
polarization with respect to the spin-valley operator. We analyze spin and
spin-valley currents and possible superconductivity in these systems. We
show that spin or spin-valley currents can flow in both half-metallic
phases.
\end{abstract}


\date{\today}

\maketitle

\section{Introduction}
\label{Intro}

Electron states at the Fermi surface of usual metals are degenerate with
respect to the spin projection. Consequently, the spin polarization of such
electron systems is zero. However, strong electron-electron interactions
can lift this degeneracy and thus, the electron liquid at the Fermi surface
acquires spin polarization. In the most extreme case, electrons with only
one spin projection (spin-up or spin-down) reach the Fermi surface, while
the states with opposite spin projection are pushed away from the Fermi
energy. These systems are referred to as
half-metals~\cite{first_half_met1983,half_met_review2008,hu2012half}.
The most immediate consequence of the half-metallicity is the perfect spin
polarization of the electric current. This makes half-metals promising
materials for applications in
spintronics~\cite{review_spintronics2004,hu2012half}.
Many rather different materials are now classified as half-metals; for example:
NiMnSb,~\cite{nimnsb_exp1990}
La$_{0.7}$Sr$_{0.3}$MnO$_{3}$,~\cite{lasrmno_half_met_exp1998}
CrO$_2$,~\cite{cro2_half_met_exp2001}
Co$_2$MnSi,~\cite{co2mnsi_half_met_exp2014}
among others. Along with the listed above ferromagnets, the half-metallicity can exist in the systems with different magnetic ordering. In Ref.~\onlinecite{M. P. Ghimire} using the first-principles density functional approach, it was shown that in double-perovskite structure [Pr$_{2-x}$Sr$_x$MgIrO$_6$]$_2$ synthesized recently, half-metal antiferromagnetism or ferrimagnetism can be observed depending on the Sr doping level.

It is commonly
accepted~\cite{half_met_review2008}
that the half-metallicity of the compounds listed above is related to an
appreciable electron-electron interaction, associated with the
transition-metal atoms. However, in recent years, transition-metal-free
half-metallicity has been a subject of intense research activity. As a
specific example, one can mention density-functional
studies~\cite{metal_free_hm2012,meta_free_hm2014},
which predict the existence of half-metallicity in graphitic carbon nitride
$g$-C$_4$N$_3$. Another well-known suggestion is to look for half-metallicity at the zigzag edges of graphene nanoribbons~\cite{son2006half}.
Some other proposals have also been discussed~\cite{kan2012half,huang2010intrinsic}.
Transition-metal-free half-metals could be of interest for bio-compatible
applications and, in general, are consistent with current interest in
carbon-based and organic-based mesoscopic
systems~\cite{soriano2010,plastic_electr2010,Avouris2007,meso_review,
chinese_phys_silicene2014,bilayer_review2016}.
The spin-orbit coupling produces a significant effect on the spin polarization and, consequently, on the condition under which the half-metallicity is observed. In the materials without transition metals, this coupling is small. In our consideration, we neglect spin-orbit interaction since the main idea of our proposal is to demonstrate that the half-metallic state can exist in the systems consisting only of light atoms, when all effects related to heavy atoms are disregarded.

A strong electron-electron interaction is not characteristic of materials
composed entirely of $s$- and $p$-elements. Therefore, it is reasonable to
focus the search for transition-metal-free half-metals on systems, in which
the electrons at the Fermi surface can be completely polarized under
the condition of weak electron-electron coupling.

In our recent work,
Ref.~\onlinecite{PhysRevLett.119.107601},
we have proposed a mechanism for half-metallicity in the weak-coupling
regime. We demonstrated that doping a spin-density wave (SDW) or
charge-density wave (CDW) insulator may stabilize a certain type of
half-metallic state provided that the undoped system has two nested
spin-degenerate Fermi surface sheets, which we will also refer to as
valleys. The nesting between the electron and hole Fermi surface sheets
makes the system unstable with respect to density wave
formation~\cite{PhysRevLett.119.107601}.
The SDW or CDW instability opens a gap, giving rise to an insulating ground
state. When doping is introduced, the system becomes metallic, with two new
Fermi surface
sheets~\cite{PhysRevLett.119.107601}.
Both sheets are half-metallic. If the spin
polarizations of the sheets are parallel to each other, a half-metallic
state, denoted below as a CDW half-metal, emerges. For antiparallel
polarizations, a different half-metallic state, the SDW or spin-valley
half-metal,
appears~\cite{PhysRevLett.119.107601}.

In this paper, we present a more detailed analysis of the previously
proposed
approach~\cite{PhysRevLett.119.107601}
to half-metallicity. The most immediate consequences of the
half-metallicity are also discussed. Specifically, we calculate the phase
diagram of the model as a function of doping. Then, the relation of the
electric current to the spin and spin-valley currents is discussed. Namely,
below we show that, depending on the specific parameters, the current
carries, in addition to the electric charge, either spin or spin-valley
quantum numbers. Finally, the structure of a possible superconducting order
parameter is discussed. Since there is no spin degeneracy in a half-metal,
but two valleys are available, the superconductivity in such a system is
rather different from that of common $s$-wave superconductors.

This paper is organized as follows. In
Sec.~\ref{sec::Model}
we formulate the model, derive its mean field solution, and construct the
model's phase diagram. Both commensurate and incommensurate density wave
order parameters are investigated. In
Sec.~\ref{sec::SpinH}
the conductivity of the system is analyzed. Superconductivity is considered
in
Sec.~\ref{sec::SC}.
Finally, the main results are discussed in
Sec.~\ref{sec::discussion}.

\section{Model}
\label{sec::Model}

\begin{figure}[t]
\centering
\includegraphics[width=0.98\columnwidth]{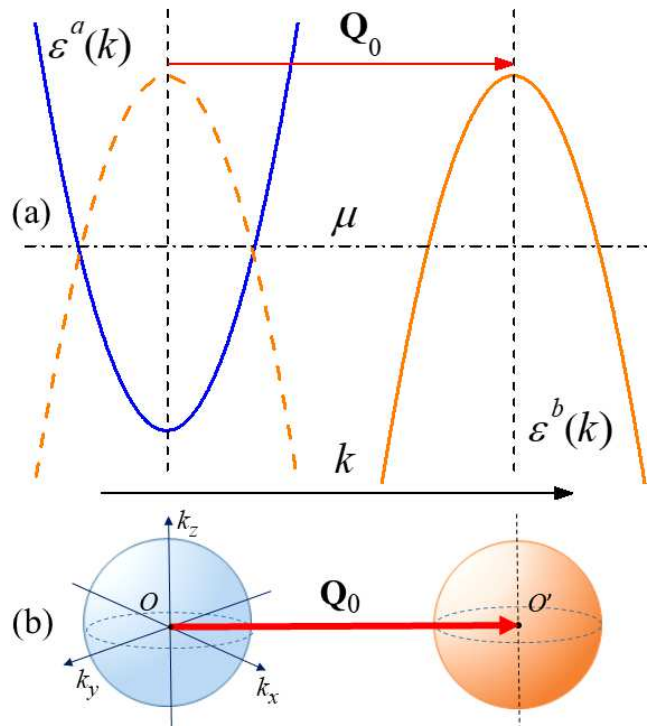}
\caption{Electron bands of the model when the electron-electron coupling is
neglected and doping is zero.
(a): Electron band
$\varepsilon^a({\bf k})$
and hole band
$\varepsilon^b({\bf k})$
are shown by solid curves. The dashed parabola is the hole band translated
by the nesting vector
${\bf Q}_0$.
The vertical axis is energy and the horizontal axis is momentum, while the
Fermi level $\mu$ is shown by the horizontal dash-dot line.
(b): Spherical Fermi surfaces of the electron and hole bands. The spheres
coincide if we translate one of them by the nesting vector.
\label{valleys1}
}
\end{figure}

We consider here an isotropic two-band electron model. Both bands or
valleys have a quadratic dispersion law. Bands $a$ and $b$ are the electron
and hole bands, respectively. The bands are schematically shown as blue and
orange parabolas in
Fig.~\ref{valleys1}(a).
Thus, the single-particle dispersions of the bands can be written as
($\hbar = 1$)
\begin{eqnarray}
\label{spectrum_a}
\varepsilon^a(\mathbf{k})=\frac{\mathbf{k}^2}{2m_a} +
\varepsilon^{a}_{\textrm{min}}-\mu,
\quad
\varepsilon^{a}_{\textrm{min}}<\varepsilon^a<\varepsilon^{a}_{\textrm{max}},
\\
\label{spectrum_b}
\varepsilon^b(\mathbf{k}\!+\!\mathbf{Q}_0) = -\frac{\mathbf{k}^2}{2m_b}
+\varepsilon^{b}_{\textrm{max}}-\mu,
\quad
\varepsilon^{b}_{\textrm{min}} < \varepsilon^b<\varepsilon^{b}_{\textrm{max}}.
\end{eqnarray}
Here, band $a$ is centered at
${\bf k} = 0$,
and band $b$ is shifted by some momentum
${\bf Q}_0$.
Below, for simplicity, we assume perfect electron-hole symmetry:
$m_a = m_b = m$
and
$\varepsilon^{b}_{\textrm{max}}
=
-\varepsilon^{a}_{\textrm{min}} = \varepsilon_{\rm F}$.
Zero doping corresponds to
$\mu = 0$.
In the absence of doping, the Fermi surface sheets for the $a$ and $b$
bands are spheres [see
Fig.~\ref{valleys1}(b)]
with the same Fermi momentum
$k_{\rm F} = \sqrt{2m \varepsilon_{\rm F}}$
and the same density of states (per spin projection)
$N_{\rm F} = {m k_{\rm F}}/(2\pi^2)$
at the Fermi energy. A model of this kind was introduced  long ago by Rice
in connection to the incommensurate SDW in
chromium~\cite{Rice}.
Hereafter,
$\varepsilon_{\rm F}$,
$k_{\rm F}$,
and
$N_{\rm F}$
denote the corresponding values at zero doping.

The quasiparticle dispersion given by
Eqs.~(\ref{spectrum_a})
and~(\ref{spectrum_b})
exhibits perfect nesting; that is, after translating the electron Fermi
surface by the vector
$\mathbf{Q}_0$,
the electron sheet completely coincides with the hole sheet, see
Fig.~\ref{valleys1}.
The vector
$\mathbf{Q}_0$
is usually referred to as the nesting vector.

In general, electrons interact with each other, so the total Hamiltonian of
the system is
\begin{equation}
\label{ham_summa}
\hat{H}=\hat{H}_e+\hat{H}_{\textrm{int}}\,.
\end{equation}
Here
$\hat{H}_e$
is the one-electron term, which corresponds to the dispersion
laws~(\ref{spectrum_a})
and~(\ref{spectrum_b}).
The term
$\hat{H}_{\textrm{int}}$
describes the interaction between quasiparticles.

We are interested in the weak-coupling regime, as it was mentioned above.
We assume that the interband and intraband interactions are of the same
order. Thus, to treat the SDW or CDW instability, it is sufficient to keep
in
$H_{\rm int}$
only the interaction between the electrons in band $a$ and holes in band
$b$,
respectively~\cite{Rice,PhysRevLett.119.107601}.
It is this term in the interaction Hamiltonian, which generates the gap and
cannot be treated as a perturbation. A weak intraband coupling can be
considered perturbatively, but this can be safely neglected because it only
provides small corrections to our results. This common feature of BCS-like
approaches can be proved by a direct calculation.

Below we assume that the interaction is a short-range one. In this case,
$\hat{H}_{\textrm{int}}$
can be written as
\begin{equation}
\label{Hint}
\hat{H}_{\textrm{int}}= \hat{H}_{\textrm{dir}}+\hat{H}_{\textrm{ex}},
\end{equation}
where
\begin{equation}
\label{eq::ham_int}
\hat{H}_{\textrm{dir}}
\!=\!
g \int\!{d^3{\bf r} \,\sum_{\sigma\sigma'}	
	\psi^\dag_{a\sigma}(\mathbf{r})	\,
	\psi^{\vphantom{\dagger}}_{a\sigma}(\mathbf{r})\,
	\psi^\dag_{b\sigma'}\!(\mathbf{r})\,
	\psi^{\vphantom{\dagger}}_{b\sigma'}\!(\mathbf{r})}\,,
\end{equation}
and
\begin{equation}
\label{eq::ham_ex}
\hat{H}_{\textrm{ex}}
\!=\!
g_\perp \int\! d^3\mathbf{r}\,
	\sum_{\sigma\sigma'}
		\psi^\dag_{a\sigma}\!(\mathbf{r})
		\psi_{b\sigma}^{\vphantom{\dagger}}(\mathbf{r})
		\psi^\dag_{b\sigma'}\!(\mathbf{r})
		\psi_{a\sigma'}^{\vphantom{\dagger}}(\mathbf{r})\,.
\end{equation}
Here,
$\psi_{\alpha \sigma}\!(\mathbf{r})$
denotes the usual fermionic field operator for band
$\alpha$\,($=a,b$)
and spin projection $\sigma$ onto the $z$ axis; and
${\bf r}$
refers to spatial coordinates. The term
$\hat{H}_{\textrm{dir}}$
represents the direct part of the density-density interaction, while
$\hat{H}_{\textrm{ex}}$
corresponds to the exchange part of this interaction. The constants $g$ and
$g_\perp$
describe the electron-hole interaction. We assume that the interaction is
repulsive
($g,\,g_\perp >0$)
and weak
($g N_{\rm F}, \,g_\perp N_{\rm F} \ll 1$).

\subsection{SDW instability and spin-valley half-metal}
\label{SDW}

Hamiltonian~\eqref{ham_summa}
can be used to describe the spontaneous formation of low-temperature
density-wave order when the Fermi surface sheets of holes and electrons
perfectly match each other (perfect nesting). We start with the SDW.
Looking ahead, we can state that the SDW order has a lower free energy than
the CDW one if we take into account only electron-electron coupling
Eqs.~\eqref{eq::ham_int}
and~\eqref{eq::ham_ex},
and disregard, say, electron-lattice interactions. Up to rotations of the
spin-polarization axis, the SDW ground state is believed to be unique. In
the weak-coupling regime, it is well described by a mean-field BCS-like
theory.

To construct a mean-field theory of the SDW order, we group the electron operators into two sectors, labeled by the spin index
$\sigma=\pm 1/2$
(or
$\sigma=\uparrow,\, \sigma=\downarrow$): sector
$\sigma$
consists of
$\psi_{a \sigma}$
and
$\psi_{b {\bar \sigma}}$
(here
${\bar \sigma}$
means
$-\sigma$).
In the zero-temperature mean-field approach, the sectors are decoupled, and
the (sector-dependent) SDW order parameter is
\begin{equation}
\label{rice}
\Delta_\sigma
=
\frac{g}{V}\sum_{\mathbf{k}}
	\left\langle
		\psi^\dag_{\mathbf{k}a\sigma}\,
		\psi^{\phantom{\dag}}_{\mathbf{k}b\bar{\sigma}}
	\right\rangle\,,
\end{equation}
where $V$ is the system volume, and
$\langle \ldots \rangle$
denotes the diagonal matrix element for the ground state. The symbol
$\psi^{\vphantom{\dagger}}_{\mathbf{k}\alpha\sigma}$
is the Fourier transform of the operator
$\psi_{\alpha\sigma}^{\vphantom{\dagger}}(\mathbf{r})$,
in which the momentum
${\bf k}$
is measured from the center of the band $\alpha$. The latter convention
simplifies the notation; however, one must remember that the centers of the
band $a$ and band $b$ are separated by the nesting vector
${\bf Q}_0$.
Consequently, the order parameter
$\Delta_\sigma$
oscillates in space with a period related to the wave vector
${\bf Q}_0$.

Following a mean-field approach, it is straightforward to check that only
the direct
interaction~\eqref{eq::ham_int}
contributes to the SDW ordering. The exchange term,
Eq.~\eqref{eq::ham_ex},
cannot be expressed as a product of two bilinear combinations of the form
$\psi^\dag_{a\sigma}\, \psi^{\phantom{\dag}}_{b\bar{\sigma}}$,
which enter the definition of order parameter~\eqref{rice}. Therefore,
$\hat{H}_{\rm ex}$ can be neglected in the lowest approximation, similar to the intravalley terms. Thus, in the mean-field approximation, the model Hamiltonian can be
rewritten as
\begin{equation}
\label{H_SDW}
\hat{H}_{\textrm{SDW}}
\!=\!
\sum_{\mathbf{k}\alpha\sigma}\!
	\left[
		\varepsilon^\alpha (\mathbf{k})
		\psi_{\mathbf{k}\alpha \sigma}^{\dag}
		\psi_{\mathbf{k}\alpha \sigma}^{\vphantom{\dagger}}
		\!-\!
		\Delta_\sigma
		\psi^\dag_{\mathbf{k}\bar{\alpha} \bar{\sigma}}
		\psi_{\mathbf{k}\alpha \sigma}^{\vphantom{\dagger}}
	\!+\!
\frac{\Delta_{\sigma}^2}{g}\right],
\end{equation}
where
$\alpha=a,b$
and
$\bar{\alpha}$
means `not
$\alpha$'. The spectrum of Hamiltonian~\eqref{H_SDW} is
\begin{equation}
\label{spectrum}
E^{(1,2)}_{\mathbf{k}\sigma}=\mp\sqrt{\varepsilon^2_{\mathbf{k}}+\Delta_\sigma^2},
\end{equation}
where
$\varepsilon_{\mathbf{k}}=k^2/2m-\varepsilon_F$.

The equilibrium parameters of the system can be derived by minimizing the
grand thermodynamic potential, defined for arbitrary temperature $T$ by the
usual formula
\begin{equation}
\label{Omega}
\Omega=-T\ln{\!\left\{
	\textrm{Tr}\,\exp[{-(\hat{H}-\mu \hat{N})/T}]
	\right\}}.
\end{equation}
In this expression,
$\hat{N}$
is the operator of the total particle number and the Boltzmann constant
$k_\textrm{B} =1$.
In the mean-field approach, the  grand potential of our system is a sum
$\Omega=\sum_\sigma\Omega_\sigma$,
where the partial grand potentials
$\Omega_\sigma$
are equal to~\cite{PhysRevLett.119.107601}
\begin{eqnarray}
\label{eq::grand_pot}
\!\!\Omega_\sigma
\!=\!
\frac{\Delta_\sigma^2 V}{g}
\!-\!
\sum_{\bf k}\!
	\left[
		\mu-E_{{\bf k}\sigma}^{(1)}
		\!+\!
		\left(\mu\!-\!E_{{\bf k}\sigma}^{(2)}\right)
		\theta\!\left(\mu\!-\!E_{{\bf k}\sigma}^{(2)}\right)\!
\right].
\end{eqnarray}
The symbol
$\theta (z)$
denotes the Heaviside step function. To describe the system at finite
doping $x$ it is convenient to introduce the partial dopings
\begin{eqnarray}
\label{eq::part_dop_def}
x_\sigma = - \frac{\partial \Omega_\sigma}{\partial \mu},
\end{eqnarray}
which are the amounts of additional charge accumulated in sector
$\sigma$.
Obviously, they satisfy
\begin{eqnarray}
\label{eq::charge_cons}
x_\uparrow + x_\downarrow = x\,.
\end{eqnarray}
The order parameter
$\Delta_{\sigma}$
minimizes the grand potential,
$\Omega_\sigma (\Delta_\sigma)$:
\begin{eqnarray}
\label{eq::Delta_min}
\frac{\partial \Omega_\sigma}{\partial \Delta_\sigma} = 0.
\end{eqnarray}
Thus, to describe the system at finite doping, one has to solve the system
of
Eqs.~(\ref{eq::part_dop_def})--(\ref{eq::Delta_min}) to obtain
$\mu$
and
$\Delta_\sigma$
as functions of
$x$.
Expressions~(\ref{eq::grand_pot})--(\ref{eq::Delta_min})
are valid provided
that the state of the system remains homogeneous, and the SDW order remains
commensurate even at finite doping (see Section~\ref{Incom} and
Ref.~\onlinecite{our_chrom2013}).
Note here that different electron pockets are usually located near the
high-symmetry points of the Brillouin zone. Thus, the vector
$\mathbf{Q}_0$
is related to the underlying lattice structure and the order may be called
commensurate. At nonzero doping, we may try to optimize the energy further
by treating the translation vector,
$\mathbf{Q}_1 = \mathbf{Q}_0 + \mathbf{Q}$,
as a variational parameter, which is not directly related to the lattice
constant. Further on, such order is referred to as an incommensurate one.

Direct calculations show that, at zero doping, the sectors in the ground
state are degenerate:
$\Delta_\uparrow = \Delta_\downarrow = \Delta_0$.
The nesting is perfect and the order parameter is equal to the BCS-like
value
\begin{eqnarray}
\Delta_0 \approx \varepsilon_{\rm F} \exp \left( -1/g N_{\rm F} \right).
\end{eqnarray}
The obvious BCS structure of this expression is a consequence of the fact
that in each sector, the mean-field procedure is mathematically equivalent
to the BCS calculations.

\begin{figure}[t]
\centering
\includegraphics[width=0.75\columnwidth]{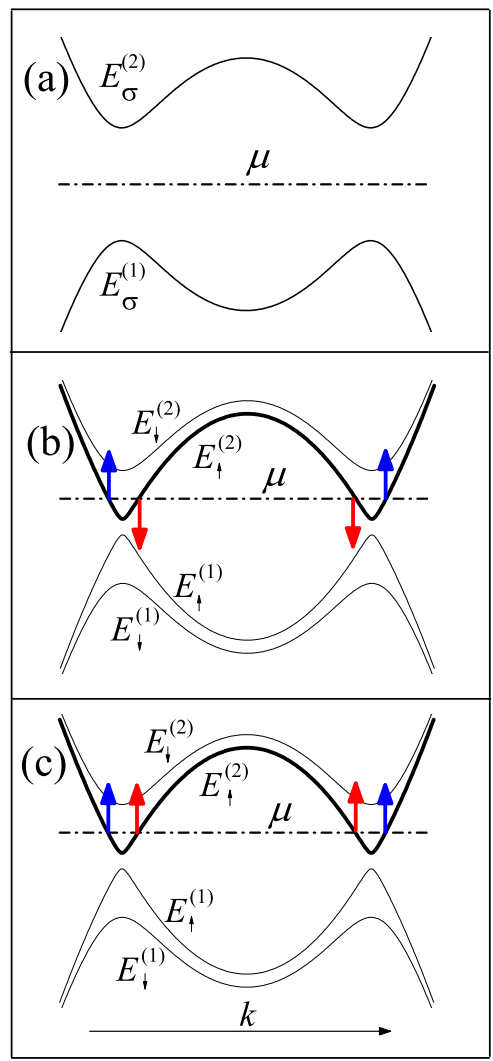}
\caption{Electron band structure for the insulating and half-metallic
states. The vertical axis is the energy, while the horizontal axis is the
momentum. The Fermi level $\mu$ is shown by the horizontal dash-dot lines.
(a): If doping is zero ($x=0$), the ground state is an insulating SDW or CDW depending on the model parameters, with degenerate sectors
($\Delta_\uparrow\equiv\Delta_\downarrow$).
The energies of electron and hole bands $E_\sigma^{(1,2)}$
are given by Eq.~\eqref{spectrum}.
(b) and~(c): If $x>0$, the sectors are no longer degenerate
($\Delta_\uparrow < \mu < \Delta_\downarrow \equiv \Delta_0$),
with the charge accumulating in sector $\uparrow$, in which a Fermi surface
appears. The spin polarizations (arrows) of the Fermi surface sheets in (b)
correspond to the spin-valley half-metal, and in (c) -- to the CDW
half-metal.
\label{valleys}
}
\end{figure}

Once
$\Delta_0$
is known, the spectrum of the model at
$x=0$
can be evaluated, see
Fig.~\ref{valleys}(a).
Note also that at zero doping, the definition of the order parameter
Eq.~\eqref{rice} implies that the total SDW polarization in real space is directed along the $x$
axis~\cite{PhysRevLett.119.107601}
\begin{eqnarray}
\label{eq::sx}
\langle S^x(\mathbf{r})\rangle
&=&
\frac{\Delta_\uparrow+\Delta_\downarrow}{2g}
	\exp (i{\bf Q}_0 {\bf r})
+ {\rm c.c.}
\\
\nonumber
&=&
\frac{2\Delta_0}{g}\cos({\bf Q}_0 {\bf r}),\\
\label{eq::sy}
\langle S^y(\mathbf{r})\rangle
&=&
\frac{\Delta_\uparrow - \Delta_\downarrow}{2ig}
\exp(i{\bf Q}_0 {\bf r}) + {\rm c.c.}
\equiv 0\,.
\end{eqnarray}

The doping shifts the chemical potential from zero and suppresses the
perfect nesting. The number of low-energy states competing to become the
true ground state increases. Both incommensurate and inhomogeneous
phases~\cite{Rice,tokatly1992,eremin_chub2010,PrlOur,PrbOur,PrbROur,
Sboychakov_PRB2013_PS_pnict,PrbVOur,prb_sl2017}
were considered as ground states of 
Hamiltonian~(\ref{ham_summa})
and its modifications. In our previous paper
Ref.~\onlinecite{PhysRevLett.119.107601},
we show that the half-metallic state is yet another viable contender in the
case of imperfect nesting. Here, we consider this problem in more detail.

According to Eqs.~(\ref{eq::part_dop_def}) and~(\ref{eq::Delta_min}), the two sectors $\sigma$ are decoupled within the mean-field approach. Then, applying a well-known procedure~\cite{tokatly1992,our_chrom2013,prb_sl2017},
one can calculate the order parameters $\Delta_\sigma$ and the chemical
potential $\mu$ as functions of $x_\sigma$. This gives the following
expression
\begin{eqnarray}
\label{eq::doped_sdw_delta}
\Delta_\sigma
=
\Delta_0 \sqrt{1 - \frac{x_\sigma}{N_{\rm F} \Delta_0}}\,,
\qquad
\mu = \Delta_0 - \frac{x_\sigma}{2 N_{\rm F}}\,.
\end{eqnarray}
We see that the doping of sector $\sigma$ destroys the order parameter in
this sector. In the homogeneous commensurate state,
$\Delta_\sigma$
is zero when
$x_\sigma \geq x_0$,
where
\begin{eqnarray}
\label{eq::x0_def}
x_0 = N_{\rm F} \Delta_0
\end{eqnarray}
is a characteristic doping level.

It is usually assumed without extra examination (see, e.g.,
Refs.~\onlinecite{Rice,our_chrom2013,prb_sl2017})
that the charge carriers are spread evenly between both sectors, that is,
\begin{eqnarray}
\label{eq::degen}
x_\uparrow = x_\downarrow = \frac{x}{2},
\quad
\text{and}
\quad
\Delta_\uparrow = \Delta_\downarrow.
\end{eqnarray}
Nevertheless, it is easy to show that the spontaneous lifting of the
degeneracy~(\ref{eq::degen})
optimizes the energy. To prove this, the system free energy $F$ must be
obtained. (Switching from $\Omega$ to $F$ is necessary to work at fixed
doping.) The free energy equals to the sum
$F=\sum_\sigma F_\sigma$,
where the partial free energy,
\begin{eqnarray}
F_{\sigma} (x_{\sigma})
=
\Omega_{\sigma} (\mu (x_\sigma)) + \mu (x_\sigma) x_\sigma
\end{eqnarray}
can be calculated as
\begin{eqnarray}
\label{FreeEn}
F_\sigma (x_\sigma)
=
F_\sigma (0) + \int\limits_0^{\,x_\sigma}\!\!dx'\mu (x')\,,
\end{eqnarray}
where
\begin{eqnarray}
F_\sigma (0) = -\frac{1}{2} N_{\rm F} \Delta^2_0
\end{eqnarray}
is a well-known BCS-like expression for the free energy at perfect nesting.
Then, using $\mu$ from
Eq.~\eqref{eq::doped_sdw_delta}, we derive
\begin{eqnarray}
\label{F_sigma}
\frac{F_\sigma}{V}
&=&
- \frac{N_{\rm F} \Delta^2_0}{2}
+
\Delta_0x_\sigma-\frac{x_\sigma^2}{4N_{\rm F}}\,,\;\;x_{\sigma}<x_0\,,
\\
\label{eq::Ftotal}
\frac{F}{V}
&=&
\sum_\sigma \frac{F_\sigma}{V}
=
-N_{\rm F} \Delta_0^2 + \Delta_0 x-\frac{x_\uparrow^2
+ x_\downarrow^2}{4 N_{\rm F}}\,.
\end{eqnarray}
Thus, only the third term in
Eq.~(\ref{eq::Ftotal}) depends on the distribution of the charge between
the two sectors. Expression~(\ref{eq::Ftotal})
has to be minimized under the
constraint~(\ref{eq::charge_cons}).
It is easy to check that $F$ has the smallest value when
$x_\sigma = x$
and
$x_{\bar \sigma}=0$.
In other words, for fixed $x$, within the studied class of spatially
homogeneous mean-field states, the most stable one corresponds to the case
when all the doped charge is accumulated in one sector. The other sector is
completely free of extra charge carriers. Thus, the degeneracy between
sectors
$\sigma=\uparrow$
and
$\sigma=\downarrow$
is lifted, and
equations~(\ref{eq::degen})
are no longer valid. To be specific, let us assume that
$\sigma=\uparrow$
represents the sector accumulating extra charge. Therefore, in the ground
state, we have
\begin{eqnarray}
\label{eq::F_half}
\frac{F}{V}&=&-N_{\rm F} \Delta_0^2 + \Delta_0 x-\frac{x^2}{4 N_{\rm F}}\,,
\\
\label{eq::mu_half}
\mu &=& \Delta_0 - \frac{x}{2N_{\rm F}}\,,
\\
\label{eq::delta_half}
\Delta_\uparrow (x)&=&\Delta_0 \sqrt{1 - \frac{x}{N_{\rm F} \Delta_0}},
\quad
\Delta_{\downarrow}(x) = \Delta_0\,.
\end{eqnarray}
These relations are valid for low doping
$x<x_0$.

An important feature of
Eq.~(\ref{eq::F_half}) is that the second derivative
$\partial^2 F/\partial x^2$
is negative. This means that the doped system may be unstable with respect
to electronic phase
separation~\cite{tokatly1992,moreo1999,dagotto_book,dagotto_phasep2003,
PrbROur,Sboychakov_PRB2013_PS_pnict,our_chrom2013,IrkhinPRB2010}.
However, the long-range Coulomb interaction can suppress phase
separation~\cite{di_castro1,bianconi2015intrinsic}.
Thus, it is reasonable to study here the properties of the homogeneous state.

It follows from
Eqs.~\eqref{eq::mu_half}
and~\eqref{eq::delta_half}
that
\begin{eqnarray}
\Delta_\uparrow(x) < \mu (x)<\Delta_{\downarrow}(x) = \Delta_0,
\,\,
\text{when}
\,\,
0<x<x_0.
\end{eqnarray}
This means that in the sector
$\downarrow$,
the order parameter remains equal to
$\Delta_0$.
Since the chemical potential is lower than
$\Delta_{\downarrow}$,
no charge enters sector
$\downarrow$,
see
Fig.~\ref{valleys}(b).
In the sector $\uparrow$, two Fermi surface sheets emerge. According to
Eqs.~\eqref{spectrum}, \eqref{eq::x0_def}, \eqref{eq::mu_half},
and~\eqref{eq::delta_half},
they are determined by
\begin{eqnarray}
\label{eq::Fermi_surface}
\!\!\!\!\!\!\!\varepsilon^2_{\bf k} = [\mu (x)]^2\! -\! [\Delta_\uparrow(x)]^2,
\,\,
\text{or}
\,\,
k = k_{\rm F}\sqrt{1\pm \frac{\Delta_0}{2\varepsilon_{\rm F}}\frac{x}{x_0}}.
\end{eqnarray}

The doped state acquires non-trivial macroscopic quantum numbers, since
charge carriers introduced by the doping are distributed unevenly between
the sectors. To characterize the macroscopic state, it is useful to specify
the spin operator
${\hat S}$
and spin-valley operator
${\hat S}_{\rm v}$:
\begin{eqnarray}
\label{operators}
{\hat S}
=
\sum_{\alpha\sigma}
	\sigma {\hat N}_{\alpha\sigma},
\quad
{\hat S}_{\rm v}
=
\sum_{\alpha\sigma} \sigma \nu_\alpha {\hat N}_{\alpha \sigma},
\end{eqnarray}
where
\begin{eqnarray}
\label{eq::number_oper}
{\hat N}_{\alpha\sigma}
=
\sum_{\bf k}
	\psi^\dag_{{\bf k}\alpha \sigma}\,
	\psi^{\vphantom{\dagger}}_{{\bf k}\alpha\sigma}.
\end{eqnarray}
Here, the operator
${\hat N}_{\alpha\sigma}$
describes the number of electrons with spin $\sigma$ in valley $\alpha$.
The index
$\nu_\alpha$
is defined according to the rule
$\nu_a = 1$,
$\nu_b = -1$.

Hamiltonian~(\ref{ham_summa}),
as well as the mean-field
Hamiltonian~\eqref{H_SDW},
commutes with both
${\hat S}$
and
${\hat S}_{\rm v}$.
The field operators satisfy obvious commutation rules
\begin{equation}
\label{commute}
[{\hat S}, \psi_{\alpha \sigma}]
=
\sigma \psi_{\alpha \sigma},
\qquad
[{\hat S}_{\rm v}, \psi_{\alpha \sigma}]
=
\sigma \nu_\alpha \psi_{\alpha \sigma}.
\end{equation}
Namely, in addition to the usual spin-projection quantum number
$\sigma$,
the field
$\psi_{\alpha\sigma}$
can be characterized by the spin-valley projection
$\sigma \nu_\alpha$.

Using
Eqs.~\eqref{commute}, it is easy to check that in the sector
$\sigma$,
both
$\psi_{a \sigma}$
and
$\psi_{b {\bar\sigma}}$
carry the same spin-valley quantum number equal to
$+\sigma$.
In the sector
${\bar \sigma}$,
the field operators correspond to a
$-\sigma$
quantum of
${\hat S}_{\rm v}$.
That is, the Fermi surface sheet of the doped system is characterized by
only one projection of the spin-valley operator. The Fermi surface sheets
with the opposite projection of
${\hat S}_{\rm v}$
are absent, since the
sector
$\sigma=\downarrow$
is gapped. Thus, the doped system can be
referred to as \textit{a spin-valley
half-metal}~\cite{PhysRevLett.119.107601}: like a classical half-metal, our
system exhibits complete polarization of the Fermi surface. However, in
contrast to the usual half-metal, the polarization is not just the spin
polarization, but rather, the spin-valley one. Therefore, the electric
current flowing through the spin-valley half-metal is completely
spin-valley polarized.

What does Fermi surface polarization of this type mean? Imagine that the
spin-valley half-metal is in the state with spin-valley projection +1.
Therefore, electron states at the Fermi energy have spin projection
$\uparrow$, hole states have $\downarrow$ projection (of course, if an
electric current is present, it is carried by electrons with spin
$\uparrow$ and holes with spin $\downarrow$).

Experimental measurements of the spin-valley polarization are likely to be
more complicated than the measurements of pure spin polarization. Indeed, to
extract the spin-valley data, it is necessary to determine how spin
polarization is distributed over the Brillouin zone, as the definition of
$S_{\rm v}$,
Eq.~(\ref{operators}),
implies. On the other hand, the spin-valley polarization may be useful for
valley filtering: if we insert perfectly spin-polarized electrical current
into a spin-valley half-metal, we can determine which valley is
participating in the transport. For example, if the current spin
polarization is $\uparrow$, it is carried by the electron valley (no holes with
$\sigma = \uparrow$
are present at the Fermi level).

A spin-valley half-metal has some similarities with the antiferromagnetic
half-metals widely discussed mostly in theoretical papers, see, e.g.,
Refs.~\onlinecite{hu2012half,M. P. Ghimire}.
In an antiferromagnetic half-metal, itinerant charge carriers at the Fermi
level are still spin polarized. However, in contrast to the usual
ferromagnetic half-metal, the magnetic moment per unit cell is zero owing
to the presence
of electrons in different bands, which compensates the spin polarization of
the itinerant electrons. In the spin-valley half-metal, we also have spin
compensation of two groups of charge carriers, but here both electron-like
and hole-like charge carriers are itinerant ones and contribute to the
Fermi energy belonging to different Fermi surface sheets.

Since the sector
$\downarrow$
is free of electrons introduced by the doping, the average values of
$\hat N_{a\downarrow}$
and
$\hat N_{b\uparrow}$
remain unaffected by the doping, while
$\langle\hat N_{a\uparrow}\rangle$
and
$\langle\hat N_{b\downarrow}\rangle$
change. Let us denote the average occupation numbers
$\langle{\hat N}_{\alpha\sigma}\rangle$
as
$N_{\alpha\sigma} =\langle{\hat N}_{\alpha\sigma}\rangle$.
It is convenient to assume that in the undoped state
$N_{\alpha\sigma}=0$.
Therefore, we can write
\begin{eqnarray}
N_{a\downarrow}=N_{b\uparrow}=0,
\quad
\text{and}
\quad
N_{a\uparrow}+N_{b\downarrow}=xV.
\end{eqnarray}
Consequently,
$S_{\rm v} = \langle {\hat S}_{\rm v} \rangle$
is proportional to
$x$
\begin{eqnarray}
S_{\rm v} = \sigma x V.
\end{eqnarray}
In a system with perfect electron-hole symmetry, we have
\begin{eqnarray}
N_{a\uparrow}=N_{b\downarrow} = \frac{x V}{2},
\end{eqnarray}
which corresponds to
$S = \langle{\hat S}\rangle\equiv0$,
for any $x$. If the symmetry is absent, then
\begin{eqnarray}
|S|\propto x.
\end{eqnarray}
However, the net spin polarization of the spin-valley half-metal meets the
inequality
\begin{eqnarray}
|S| < |S_{\rm v}|.
\end{eqnarray}

The doping also affects the SDW order inherited from the undoped state. Intuitively, since the charge is accumulated only in one of the two sectors, the order parameters in different sectors become unequal to each other for
$x>0$
[Eqs.~(\ref{eq::delta_half})
express this fact mathematically]. As a result, the simple SDW is replaced
by a more complicated order parameter. Analyzing
Eqs.~\eqref{eq::sx}
and~\eqref{eq::sy},
one can prove that at finite doping, a circularly polarized spin component
emerges
\begin{equation}
\label{eq::sdw_circ}
\delta {\bf S} ({\bf r})
=
\left(
	\begin{matrix}
		\delta S^x(\mathbf{r})\\
		\delta S^y(\mathbf{r})
	\end{matrix}
\right)
=
\frac{\Delta_\uparrow\! -\! \Delta_\downarrow }{g}
\left(
	\begin{matrix}
		-\!\cos ({\bf Q}_0 {\bf r})\\
		\sin({\bf Q}_0 {\bf r})
	\end{matrix}
\right).
\end{equation}
The amplitude of this component grows as
$1-\sqrt{1-x/x_0}$,
when the doping increases.

\begin{figure}[t]
\centering
\includegraphics[width=0.99\columnwidth]{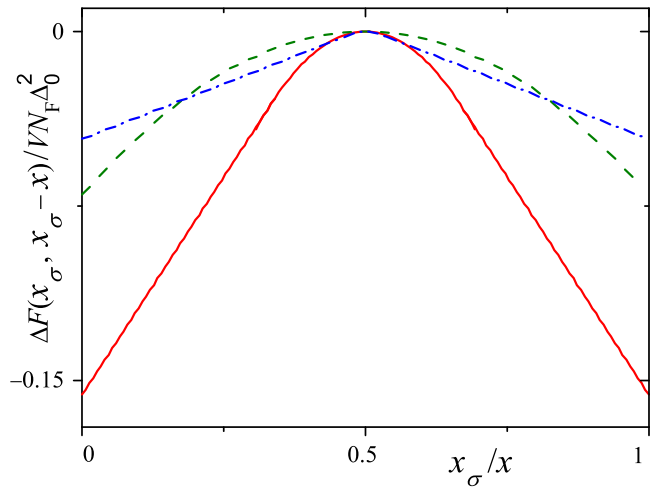}
\caption{Dependence of
$\Delta F=F_\sigma(x_\sigma)+F_\sigma(x-x_\sigma)-2F(x/2)$
on the partial doping
$x_\sigma$
calculated at
$T=0$
and fixed total doping:
$x=0.75x_0$
[(green)
dashed curve],
$x=1.5 x_0$
[(red) solid curve], and
$x=1.9x_0$
[(blue)
dash-dot curve]. The free energy curves for all three doping values have a
global maximum at
$x_\sigma = x/2$,
implying that the usual metallic phase is unstable. The free energy is the
lowest for either
$x_\sigma = 0$
or
$x_\sigma = x$:
the free energy minimum at
$x_\sigma = 0$
($x_\sigma = x$)
represents a half-metallic state with empty (filled) sector $\sigma$ and
filled (empty) sector
$\bar{\sigma}$.
\label{FigFreeEnergyCom}
}
\end{figure}

The above considerations are valid if the doping
$x$
is less than
$x_0$.
To investigate the behavior of the system in a wider doping range, we calculate the function
\begin{eqnarray}
\label{DeltaFreeCom}
\Delta F(x,x_\sigma)
=
F_\sigma(x_\sigma) + F_\sigma(x-x_\sigma)-2F_\sigma(x/2)\,.
\end{eqnarray}
If
$x<x_0$,
the doping in both sectors is less than
$x_0$.
In this case, the free energy
$F_\sigma(x_\sigma)$
is determined by
Eq.~\eqref{F_sigma} and
\begin{equation}
\label{DeltaF_Com1}
\frac{\Delta F(x,x_\sigma)}{V}
=
\frac{1}{N_{\textrm{F}}}
\left[
	-\frac{x^2}{8}+\frac{x_\sigma(x-x_\sigma)}{2}
\right].
\end{equation}
The corresponding parabolic curve is shown in
Fig.~\ref{FigFreeEnergyCom}
for
$x=0.75x_0$
by a dashed line as a function of the ratio
$x_\sigma/x$.
This function is negative and reaches its minimum when all charge carriers
introduced by the doping are concentrated within one sector (that is, when
either
$x_\sigma=0$,
or
$x_\sigma=x$);
whereas the maximum of the function
$\Delta F(x,x_\sigma)$
represents the usual SDW state with
$x_\uparrow=x_\downarrow=x/2$.
This means that the ground state corresponds to the spin-valley half-metal
phase, while the usual SDW phase is unstable, in agreement with the results
obtained above.

\begin{figure}[t]
\centering
\includegraphics[width=0.99\columnwidth]{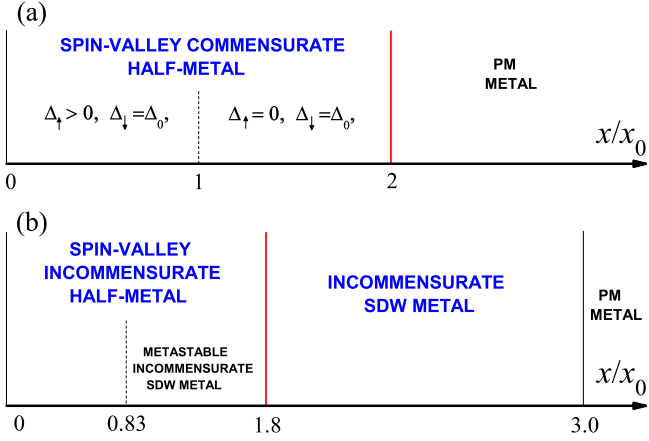}
\caption{Phase diagram of the system; (a) commensurate and (b)
incommensurate ordering. (a) Spin-valley half-metals exist within the
doping range
$0<x<2x_0$.
At
$x=x_0$
(vertical dashed line) order parameter 
$\Delta_\uparrow$
vanishes and a second-order phase transition occurs. However, a
characteristic polarization of the charge carriers at the Fermi surface
(half-metallicity) is not destroyed. When
$x=2x_0$
[vertical (red)
solid line] a first-order transition occurs from the spin-valley half-metal
phase to the PM phase.
(b) Spin-valley half-metal exists within the doping range
$0<x\lesssim
1.8x_0$.
At
$x\approx 1.8 x_0$
[(red) solid line] a second-order phase transition occurs from the
spin-valley phase to the usual SDW incommensurate state. If
$x\approx 3 x_0$
[vertical thin (black) solid line] a first-order phase transition occurs to
the PM phase. The dashed vertical line shows the point
($x\approx 0.83x_0$)
when the incommensurate SDW order can
exist as a metastable phase.
\label{PhaseDiag}
}
\end{figure}

When
$x>x_0$,
the doping in one sector can be larger than
$x_0$.
If
$x_\sigma>x_0$,
the order parameter in sector $\sigma$ vanishes, and the partial free
energy becomes
\begin{eqnarray}
F_\sigma(x_\sigma) = \frac{x_\sigma^2}{4N_{\textrm{F}}},
\end{eqnarray}
as in the disordered paramagnetic (PM) phase. Thus, for
$x>x_0$,
the function
$\Delta F(x,x_\sigma)$
is a piecewise function with the continuous first (but not second) derivative
$\partial \Delta F/\partial x_\sigma$.
In the vicinity of the point
$x_\sigma=x/2$,
the function
$\Delta F$
has a parabolic shape. It coincides with linear functions of
$x_\sigma$
away from that point, see the (red) solid
($x=1.5x_0$)
and (blue) dot-dash
($x=1.9x_0$)
curves in
Fig.~\ref{FigFreeEnergyCom}.
However, the function
$\Delta F(x,x_\sigma)$
is negative and attains a minimum if either
$x_\sigma=0$
or
$x_\sigma=x$.
Therefore, the ground state of the model is, again, a spin-valley
half-metal. In doing so, we readily obtain that a second order transition
occurs at
$x=x_0$,
where the gap in the doped sector is closed. Comparing the free energies of
the spin-valley half-metal phase and of the usual PM state, we conclude
that the PM state becomes favorable when
$x=2x_0$.
At this point, the gap in the undoped sector closes in a jump-like manner,
and a first order transition to the usual PM phase occurs. The obtained
results are summarized in
Fig.~\ref{PhaseDiag}(a).

\subsection{CDW half-metal}
\label{CDW}

The CDW order is characterized by a finite average value
$\langle \hat{\rho} ({\bf r}) \rangle$
of the density operator
\begin{eqnarray}
\hat{\rho} ({\bf r})
=
\sum_{\sigma {\bf k}}
	\psi^\dag_{\mathbf{k}a\sigma} 		
	\psi^{\phantom{\dag}}_{\mathbf{k}b\sigma}
\exp (i{\bf Q}_0 {\bf r})
+
{\rm h.c.}
\end{eqnarray}
The CDW order is described by a formalism similar to the one developed
above for the SDW. To switch between the two types of density waves, the
mean-field sectors must be redefined. Specifically, we will assume below
that the sector $\sigma$ consists of the operators
$\psi_{a \sigma}$
and
$\psi_{b \sigma}$.
This rearrangement of the sectors may be formally expressed by the substitution
\begin{eqnarray}
\label{eq::substit}
\psi_{b \uparrow}\rightarrow\psi_{b \downarrow},
\quad
\psi_{b {\downarrow}}\rightarrow\psi_{b \uparrow}.
\end{eqnarray}
Under this substitution, we have
\begin{eqnarray}
\sum_{{\bf k} \sigma}
	\langle
		\psi^\dag_{\mathbf{k}a\sigma}
		\psi^{\phantom{\dag}}_{\mathbf{k}b\bar{\sigma}}	
	\rangle
\rightarrow
\sum_{{\bf k} \sigma }
	\langle		
		\psi^\dag_{\mathbf{k}a\sigma} 		
		\psi^{\phantom{\dag}}_{\mathbf{k}b\sigma}
\rangle.
\end{eqnarray}
Therefore, the finite modulation of the spin density is replaced by a
finite modulation of the charge density:
\begin{eqnarray}
\label{rho}
2\langle {\hat S}^x ({\bf r}) \rangle
\rightarrow
\langle \hat\rho ({\bf r}) \rangle.
\end{eqnarray}
Equation~(\ref{eq::substit})
allows us to adopt the results derived for the SDW to describe the CDW
state with little modifications.

In the CDW phase, we use the finite expectation values of
$\langle
	\psi^\dag_{\mathbf{k}a\sigma}
	\psi^{\phantom{\dag}}_{\mathbf{k}b\sigma}
\rangle$
and
$\langle
	\psi^\dag_{\mathbf{k}b\sigma}
	\psi^{\phantom{\dag}}_{\mathbf{k}a\sigma}
\rangle$
to apply the mean-field decoupling in Hamiltonians~(\ref{eq::ham_int})
and~(\ref{eq::ham_ex}).
Unlike the SDW case, both the direct and exchange terms contribute to the
mean-field Hamiltonian of the CDW phase:
\begin{eqnarray}
\label{H_CDW}
\!\!\!\!\!\!\hat{H}_{\textrm{CDW}}
\!\!\!&=&\!\!\!\!
\sum_{\mathbf{k}\sigma\alpha}
	\!\left[\!
		\varepsilon^\alpha (\mathbf{k})
		\psi_{\mathbf{k}\alpha \sigma}^{\dag}
		\psi^{\phantom{\dag}}_{\mathbf{k}\alpha \sigma}
		\!-\!
		\tilde \Delta^{\vphantom{\dagger}}_\sigma
		\psi^{\dag}_{\mathbf{k}\alpha \sigma}
		\psi^{\phantom{\dag}}_{\mathbf{k}\bar{\alpha} \sigma}
		\!+\!
		\frac{\tilde{\Delta}^2_\sigma}{\tilde{g}}
	\!\right]\!\!,
\\
\label{CDWrice}
\!\!\!\!\!\!\tilde{\Delta}_\sigma
&=&
\frac{\tilde{g}}{V}
\sum_{\mathbf{k}}
	\left\langle
		\psi^\dag_{\mathbf{k}a\sigma} \,
		\psi^{\phantom{\dag}}_{\mathbf{k}b\sigma}
	\right\rangle\,,
\end{eqnarray}
where
\begin{eqnarray}\label{renorm}
\tilde{g}=g-2g_\bot
\end{eqnarray}
is the renormalized electron-electron coupling.
Hamiltonian~\eqref{H_CDW}
is similar to the SDW Hamiltonian,
Eq.~\eqref{H_SDW}.
Thus, as expected, the CDW problem is mapped onto the SDW one solved in the
previous Section. In particular, the CDW order parameter at zero doping is
$\tilde{\Delta}_0
\approx
\varepsilon_{\rm F} \exp \left( -1/\tilde{g} N_{\rm F} \right)$.
Since
$g_\bot>0$
(repulsive interaction), the CDW is always either metastable
($\tilde{\Delta}_0<\Delta_0$),
or absolutely unstable
($2g_\bot\geq g \Leftrightarrow \tilde g < 0$).
Of course, the stability of the CDW order may be improved by adding
parameters, which are beyond our simple model; for example, also considering
an applied magnetic field and the interaction with the lattice.

Calculations identical (up to relabeling) to the case of the SDW order demonstrate that for
$x > 0$
the charge carriers are accumulated in a single mean-field sector. However,
the sectors structure is changed by the
transformation~(\ref{eq::substit}):
unlike the case of spin-valley half-metals, now both electronic fields
within a single sector have the same spin projection. Therefore, if the
introduced charge fills sector $\sigma$, both Fermi surface sheets have
identical spin polarizations equal to $\sigma$, see
Fig.~\ref{valleys}(c).
This perfect polarization of the Fermi surface is a hallmark feature of
half-metals. Thus, the spin-valley half-metal is related to the CDW
half-metal by substitution~(\ref{eq::substit}).
This substitution, in particular, switches the operators
$\hat S$
and
$\hat S_{\rm v}$.
Consequently, in the CDW half-metal, we have
\begin{eqnarray}
\label{SPIN}
S = \sigma x V,
\quad
\text{and}
\quad
|S_{\rm v}| < |S|,
\end{eqnarray}
and
$S_{\rm v}=0$
in the case of the perfect electron-hole symmetry.
When
$x>0$,
in addition to the CDW order parameter, the SDW order parameter
$\langle \delta S^z \rangle$
is generated:
\begin{eqnarray}
\langle \delta S^z ({\bf r}) \rangle =
\frac{\tilde \Delta_\uparrow\! -\! \tilde \Delta_\downarrow }{g}
\cos ({\bf Q}_0 {\bf r}).
\end{eqnarray}
It grows monotonically with $x$. This is a direct analog of
Eq.~(\ref{eq::sdw_circ}).

In the case of CDWs, we obtain formulas for the free energy, chemical
potential, and order parameter similar to
Eqs.~(\ref{eq::F_half})--(\ref{eq::delta_half}),
replacing
$\Delta_0$
by
$\tilde{\Delta}_0$.
Thus, the CDW order parameter is at least metastable in the doping range
\begin{eqnarray}
0<x<2\tilde{x}_0=2N_\textrm{F}\tilde{\Delta}_0.
\end{eqnarray}
Since
$\tilde{x}_0<x_0$,
the CDW phase becomes absolutely unstable at lower doping value than that in the SDW. To illustrate this, let us now calculate the difference in the free
energy between the CDW half-metal and the spin-valley half-metal
\begin{equation}
\label{FreeDelta}
\frac{\Delta F}{V}
=
N_\textrm{F}\left(\Delta_0^2-\tilde{\Delta}_0^2\right)
-
\left(\Delta_0-\tilde{\Delta}_0\right)x.
\end{equation}
It is easy to see that, as long as
$x<\tilde{x}_0$
and
$\Delta_0>\tilde{\Delta}_0$,
the difference
$\Delta F$
decreases when doping grows; however, it is always positive. Thus, we
conclude that the spin-valley state is more stable than the CDW half-metal
phase.

\subsection{Incommensurate ordering}
\label{Incom}

Here we analyze a possible incommensurate ordering in the model under
discussion~\cite{Rice,our_chrom2013}.
We start with the SDW order. The order parameter
$\Delta_\sigma$,
calculated in the previous sections, couples electrons with unequal
momenta. Consequently, in coordinate space, the local spin polarization
rotates with wave vector
$\mathbf{Q}_0$.
Typically, the centers of different Fermi surface pockets are located near the high-symmetry  points of the Brillouin zone. Therefore, the vector
$\mathbf{Q}_0$
is related to the underlying lattice structure. Such an order may be called
commensurate. Yet, as it has been already mentioned above, we may try to
relax the requirement of the commensurability and optimize the energy
further by treating the translation vector
$\mathbf{Q}_1 = \mathbf{Q}_0 + \mathbf{Q}$
as a variational parameter. The new order parameter has the form
\begin{equation}
\label{SDW_Icom}
\Delta_\sigma(\mathbf{Q})
=
\frac{g}{V}
\sum_{\mathbf{k}}
	\left\langle
		\psi^\dag_{\mathbf{k}a\sigma} \,
		\psi^{\vphantom{\dagger}}_{\mathbf{k+Q}b\bar{\sigma}}
	\right\rangle\,,
\end{equation}
where, as before, the momentum for electrons in band $\alpha$ is measured
from the center of the band $\alpha$. The vector
$\mathbf{Q}$
is small,
\begin{eqnarray}
|\mathbf{Q}|\sim \Delta_0 m/k_\textrm{F}\ll |\mathbf{Q}_0|.
\end{eqnarray}
The order
parameter~\eqref{SDW_Icom}
describes the SDW order with a rotating spin polarization. This rotation is
characterized by the spatial period
$2\pi/|\mathbf{Q}_0+\mathbf{Q}|$.
This value is unrelated to the underlying lattice and such order is called incommensurate.

To describe the incommensurate state, we calculate the grand potential
$\Omega$.
In the mean-field approach,
$\Omega$
is a sum of grand potentials
$\Omega_\sigma = \Omega_\sigma (x_\sigma)$.
Similar to
Eq.~\eqref{spectrum},
the eigenvalues of the mean-field Hamiltonian are
\begin{equation}
\label{spectrumQ}
\!\!E^{(1,2)}_{\mathbf{k}\sigma}
\!=\!
\frac{\varepsilon_{\mathbf{k+Q}}\!-\!\varepsilon_{\mathbf{k}}}{2}
\!\mp\!
\sqrt{\Delta_\sigma^2(\mathbf{Q})
\!+\!
\left[
	\frac{\varepsilon_{\mathbf{k+Q}}\!+\!\varepsilon_{\mathbf{k}}}{2}
\right]^2}\,.
\end{equation}
With this new formula for
$E^{(1,2)}_{\mathbf{k}\sigma}$,
the expression for the partial grand potentials
$\Omega_\sigma$,
Eq.~\eqref{eq::grand_pot},
remains unchanged. We add the minimization condition
$\partial\Omega_\sigma/\partial Q=0$
to
Eqs.~(\ref{eq::part_dop_def})--(\ref{eq::Delta_min}) and solve the obtained system numerically as it was described in
Ref.~\onlinecite{our_chrom2013}
[see Eqs.~(11)--(20) of that paper].

\begin{figure}[t]
\centering
\includegraphics[width=0.99\columnwidth]{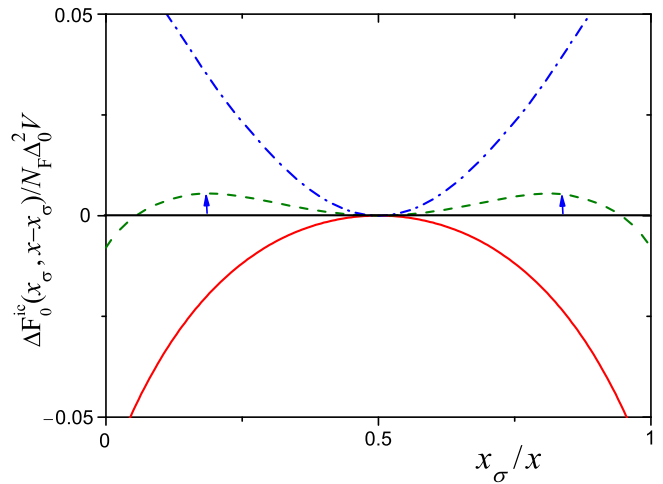}
\caption{Dependence of
$\Delta F_0^{\rm ic}(x_\sigma,x-x_\sigma)
\equiv
F_0^{\rm ic}(x_\sigma)+F_0^{\rm ic}(x-x_\sigma)-2F_0^{\rm ic}(x/2)$
on the partial doping
$x_\sigma$,
calculated at
$T=0$
and fixed total doping
$x=1.4x_0$
[(red) solid curve],
$x=1.76x_0$
[(green) dashed curve], and
$x=2.0x_0$
[(blue) dash-dot curve]. At high doping
($x=2.0x_0$),
the state at
$x_\sigma = x_{\bar \sigma} = x/2$
has the lowest free energy, therefore, the usual metal, with even
distribution of the doped charges among the sectors, is a stable phase.
When the doping is low
($x=1.4x_0$),
the half-metal is stable. In this situation, the free energy minimum at
$x_\sigma = 0$
($x_\sigma = x$)
represents a half-metallic state with empty (filled) sector $\sigma$ and
filled (empty) sector
$\bar{\sigma}$.
At some intermediate doping 
$1.4 x_0 < x^* < 2.0 x_0$,
a first order transition from the usual metal to the half-metal occurs.
Near the transition, one of the phases may become metastable.  For example,
a well-defined local (but not global) minimum of the free energy at
$x_\sigma = x_{\bar \sigma} = x/2$
is clearly seen for the (green) dashed curve. This implies that for
$x=1.76x_0$,
the usual metal is metastable, while the half-metal is truly stable. The
activation barriers for the transition into the more stable half-metallic
phase are shown by the vertical arrows. The presence of the metastable
phase is marked in
Fig.~\ref{PhaseDiag}(b).
\label{FigFreeEnergy}
}
\end{figure}

The partial free energy
$F_0^{\rm ic} (x_\sigma)$
of a sector with partial doping
$x_\sigma$
in the incommensurate state is calculated according to
Eq.~\eqref{FreeEn}. Within the considered mean-field approach, the free energy of the system in the presence of the incommensurate SDW equals
\begin{eqnarray}
F^{\rm ic} (x)
=
\min_{x_\uparrow + x_\downarrow = x}
\left[	
	F_0^{\rm ic} (x_\uparrow) + F_0^{\rm ic} (x_\downarrow)
\right]\,.
\end{eqnarray}
The free energy of the system in the ground state is found by its minimization under the condition
$x_\uparrow + x_\downarrow = x$.
Our numerical analysis shows that
\begin{eqnarray}
\frac{\partial^2 F_0^{\rm ic}(x_\sigma)}{\partial x_\sigma^2} < 0,
\end{eqnarray}
for
$x_\sigma$
less than the threshold value
$x^*\cong0.83x_0$.
Since the second derivative of
$F_0^{\rm ic}$
is negative, the sum
$F_0^{\rm ic} (x_\uparrow) + F_0^{\rm ic} (x-x_\uparrow)$
as a function of
$x_\uparrow \in [0, x]$
is concave at not too large $x$. Consequently, the extremum of the latter
sum at
$x_\uparrow = x/2$
corresponds to a maximum, not a minimum [see
Fig.~\ref{FigFreeEnergy}].
Therefore, the total free energy is minimized as follows:
\begin{eqnarray}
\label{DeltaFreeIncom}
F^{\rm ic} (x)
=
F_0^{\rm ic} (x) + F_0^{\rm ic} (0),\,\,
\textrm{at}\,\,
x_\sigma = x\,\,
\textrm{and}\,\,
x_{\bar \sigma} = 0.
\end{eqnarray}
Thus, the undoped sector
${\bar \sigma}$
remains insulating. All doped charge goes to sector $\sigma$, which becomes
metallic, with a well-defined Fermi surface, and we recover the spin-valley
half-metal with an incommensurate SDW.

Note that the compressibility of the material has the same sign as the
second derivative of its free energy. Hence, the compressibility of the
system under study is negative at low doping. This is a rather general
feature of models with imperfect nesting, which, in particular, gives rise
to the possibility of phase separation in
them~\cite{tokatly1992,PrbROur,Sboychakov_PRB2013_PS_pnict,our_chrom2013,
IrkhinPRB2010}.

If
$x_\sigma>x^*$,
then
\begin{eqnarray}
\frac{\partial^2 F_0^{\rm ic}(x_\sigma)}{\partial x_\sigma^2} > 0,
\end{eqnarray}
and the total free energy
$F_0^{\rm ic}(x_\sigma)+F_0^{\rm ic}(x-x_\sigma)$
acquires a local minimum at
$x_{\uparrow}=x_{\downarrow}=x/2$
(see
Fig.~\ref{FigFreeEnergy}).
When doping increases even further, this minimum becomes a global minimum
for
$x\cong1.8N_F\Delta_0$.
Consequently, the first order transition from incommensurate spin-valley
half-metal to the usual incommensurate SDW phase occurs at this point.

The results obtained are summarized in
Fig.~\ref{PhaseDiag}(b).
Comparing them with the case of commensurate order
[Fig.~\ref{PhaseDiag}(a)],
we observe a definite difference. While the spin-valley half-metal exists
in both cases approximately within the same doping ranges, the transition
from the half-metal to the PM phase occurs in a different way: directly
from the half-metal to the PM if
${\bf Q}=0$
and via the intermediate incommensurate SDW state if
${\bf Q} \neq 0$.

Comparing the computed free energies of the commensurate and incommensurate
phases, we see that the incommensurate phase is more stable than the
commensurate one. Accounting for the incommensurability allows us to extend
the range of existence for the ordered state, as one can notice comparing
Fig.~\ref{PhaseDiag}(a)
and~\ref{PhaseDiag}(b).
However, the difference between the ordering with
${\bf Q}=0$
and
${\bf Q} \neq 0$
is small. The contributions, which are ignored in our treatment (e.g.,
disorder), can be favorable for commensurate ordering.

The results for the CDW phase can be obtained from the above calculations
by a simple replacement
${\Delta}_0\rightarrow \tilde{\Delta}_0$,
and, consequently, the incommensurate CDW half-metal is the ground state of
the system at low doping.

Among the four mean-field states discussed here (commensurate SDW/CDW
half-metals, incommensurate SDW/CDW half-metals), the incommensurate SDW
has the lowest energy at low doping, within the framework of our model.
However, the difference in free energy between the SDW and CDW phases may
be small. Indeed, the direct interaction parameter $g$ equals to
$g (0)$,
where
$g({\bf k})$
is the Fourier transform of the inter-electron repulsion energy
$g({\bf r})$,
while the exchange interaction parameter
$g_\bot$
represents the interaction at the momentum transfer
$\mathbf{Q}_1\approx \mathbf{Q}_0$:
$g_\bot = g ({\bf Q}_1)$.
If
$g ({\bf Q}_1)\ll g(0)$
(e.g., as in the case of bare Coulomb repulsion), then,
$g_\bot\ll g$
and
$\Delta\approx \tilde{\Delta}$.
Also, other factors, which are not included in our study, could favor the
CDW half-metal. For example, the proximity to a lattice instability can
make the CDW half-metal a ground state. The applied magnetic field acts
similarly, since the total spin of the CDW half-metal exceeds the spin of
the spin-valley half-metal.

\section{Electric, spin, and spin-valley conductivities}
\label{sec::SpinH}

In the system under study, the charge carriers at the Fermi surface are
spin or spin-valley polarized. Consequently, the currents are also
polarized. The problem of the polarized currents in our half-metal deserves
a separate investigation and here we only discuss this very briefly. In
particular, we assume the perfect electron-hole symmetry and consider only
commensurate ordering, since in the case of incommensurate SDW or CDW the
results are qualitatively similar.

The electrical conductivity of the isotropic system at zero temperature in
the free-electron approximation can be written
as~\cite{han2013modern}
\begin{equation}
\label{ConductE}
\sigma_{\rm E}=\frac{e^2}{3}N_{\textrm{F}}(\mu)\,\tau(\mu)\,v^2(\mu),
\end{equation}
where
$N_{\textrm{F}}(\mu)$
is the density of states,
$\tau(\mu)$
is the mean free time, and
$v(\mu)= |\partial \varepsilon_\mathbf{k}/\partial {\bf k}|$
is the electron velocity, and all values here are taken at the Fermi level
$\mu$. For simplicity, we assume further that the mean free time is the
same for electrons and holes and is independent of $\mu$. In
Eq.~\eqref{ConductE},
electrons with both spin projections are taken into account. For a
quadratic electron dispersion, we have
$\sigma_{\rm E}=e^2n\tau/m$,
where $n$ is the electron density in the conduction band.

If we neglect the electron-hole coupling, the conductivity of the two-band
system,
Fig.~\ref{valleys1},
is the sum of the electron and hole conductivities,
$\sigma_{\rm E}=\sigma_a+\sigma_b$.
When doping is zero, we have
\begin{eqnarray}
\sigma_a=\sigma_b=e^2n_0\tau/m=\sigma_0,
\quad
\text{and thus,}
\quad
\sigma_{\rm E}=2\sigma_0.
\end{eqnarray}
Here
$n_0=k^3_\textrm{F}/(6\pi^2)$
is the density of electrons
($n_a$)
or holes
($n_b$)
in the conduction band at zero doping. If we dope the system
electronically, then,
$n_a=n_0+x$.
Assuming that
$x\ll n_0$,
we obtain in the linear approximation:
$n_b\approx n_0-x$.
Therefore, the electrical conductivity remains approximately constant,
$\sigma_\textrm{E}(x)\approx 2\sigma_0$.
In this framework, the Fermi surface is spin degenerate; consequently, the
corresponding spin conductivity is zero.

\subsection{Spin-valley half-metal}
\label{SDW_current}

First, we consider the case of SDW instability and spin-valley half-metal.
The electron-hole coupling opens a gap in the spectrum and the conductivity
in the system becomes equal to zero at zero doping.

At finite doping, the mobile charge carriers are accumulated in the
conduction bands. When
$x<x_0$,
the band corresponding to the sector
$\sigma=+1/2$
is filled, while the band corresponding to
$\sigma=-1/2$
is empty. We have two Fermi-pockets in the filled band, one electron-like
($\partial E_\mathbf{k}/\partial k>0$)
and one hole-like,
($\partial E_\mathbf{k}/\partial k<0$),
see
Fig.~\ref{valleys}.
The Fermi momenta of these pockets are given by
Eq.~\eqref{eq::Fermi_surface},
where
$\mu=E_{\mathbf{k}\sigma}$.
Using
Eqs.~\eqref{eq::doped_sdw_delta} and having in mind that
$\Delta/\varepsilon_\textrm{F}\ll 1$,
we derive
\begin{equation}
\label{pockets}
\frac{k_{\rm F}^{(e,h)}}{k_{\rm F}}= 1\pm\frac{\Delta_0}{4\varepsilon_{\rm F}}\frac{x}{x_0}\approx 1,\quad x<x_0,
\end{equation}
where the superscript $e$ (superscript $h$) and the plus (minus) sign
corresponds to the electron (hole) pocket. Recall that
$k_{\rm F}$
and
$\varepsilon_{\rm F}$
denote the corresponding values at zero doping. In the same approximation,
we have
\begin{eqnarray}
\label{speed}
v_F^{(e,h)}\!\! \approx\! \pm\frac{x}{2x_0\!-\!x}\frac{k_{F}}{m},\,\, N_F^{(e,h)}\!\!\approx\! \frac{2x_0\!-\!x}{x}N_F,\,\, x<x_0.
\end{eqnarray}
Therefore, in the lowest-order approximation, when
$\Delta_0/\varepsilon_F\ll 1$,
the electron-hole symmetry is preserved.

The energy gap
$\Delta_\sigma$
in the
$\sigma=+1/2$
sector vanishes if
$x>x_0$,
while the filling of  the
$\sigma=-1/2$
sector remains zero (see Section~\ref{SDW}). Thus, for
$\Delta_0/\varepsilon_\textrm{F}\ll 1$,
the conductivity becomes
$\sigma_0$,
that is, one-half of the conductivity of the system in the PM state.

Now we can calculate the electric conductivity
$\sigma_\textrm{E}$,
which is the sum of the electron,
$\sigma_e$,
and the hole,
$\sigma_h$,
contributions. Using
Eq.~\eqref{ConductE} and \eqref{speed}, we obtain
\begin{eqnarray}
\label{Electric_SDW}
\sigma_\textrm{E}
=
\sigma_0G(x)\approx\sigma_0\!
	\left\{\!
		\begin{array}{ll}
			x/(2x_0-x), & \hbox{$0<x<x_0$,} \\
			1, & \hbox{$x_0<x<2x_0$.}
		\end{array}
\right.
\end{eqnarray}
The derivative of the function
$\sigma_\textrm{E}(x)$
has a singularity at
$x=x_0$,
when the second order transition occurs [see
Fig.~\ref{PhaseDiag}(a)].
When
$x>2x_0$,
the half-metal phase disappears, the spin degeneracy of the Fermi surface
is restored, and the conductivity exhibits a stepwise change from
$\sigma\approx\sigma_0$
to
$\sigma\approx2\sigma_0$.

The conductivity in the half-metallic state is of the order of
$\sigma_0$
if the doping $x$ is not small,
$x\sim x_0$.
The results obtained are valid if the temperature $T$ and scattering
frequency
$1/\tau$
are both smaller than the characteristic energy
$\mu-\Delta_\sigma$,
which is necessary ``to mix'' the electrons-like and the hole-like
excitations. When
$x\sim x_0$,
this means that both
$T,1/\tau\ll \Delta_0$.

If the electric current $j$ is spin-polarized, the spin current
$j^s$
associated with $j$ is nonzero. We can define the spin current as
$j^s=j\langle s_z\rangle/e$,
where
$\langle s_z\rangle$
is the average spin projection per one electron or hole at the Fermi
surface. Using
Eq.~\eqref{ConductE},
we define a spin conductivity as
\begin{equation}
\label{ConductSp}
\sigma^s=\frac{e\langle s_z\rangle}{3}\,
N_{\textrm{F}}(\mu)\,\tau(\mu)\,v^2(\mu),
\end{equation}
where we assume that there are no magnetic impurities in the sample. The
spin conductivity of the system is the sum of spin conductivities in the
electron
$\sigma_e^s$,
and the hole,
$\sigma_h^s$
pockets. To calculate
$\sigma^s$,
we need to know the spin polarization of the Fermi surface valleys. At
small doping
$x \ll x_0$,
the valley polarizations are weak
$| \langle s_z \rangle | \ll 1/2$.
They grow as the doping increases, and saturate when
$x\thicksim x_0$.
In this case, we have
$\langle s_z\rangle\approx 1/2$
for the electrons and
$\langle s_z\rangle\approx -1/2$
for holes. Therefore,
$\sigma_e^s\approx -\sigma_h^s$,
and
\begin{eqnarray}
\label{Spin_condSDW}
\sigma^s \approx 0
\end{eqnarray}
with an accuracy
$\sim\sigma^s_0(\Delta_0/\varepsilon_{\rm F})$,
where
$\sigma^s_0=\sigma_0/2e$.

In our system, we can define the spin-valley conductivity as well. Indeed,
similarly to the electron spin, we can attribute the spin-valley quantum
number
$\pm 1/2$
to the electron states at the Fermi energy, see
Eq.~\eqref{commute}.
When the electrical current flows through the system, it can carry this
quantum number, in addition to the charge.

To specify the spin-valley conductivity, we replace the spin polarization
$\langle s_z\rangle$
in
Eq.~\eqref{ConductSp}
by an average spin-valley projection
$\langle s_v\rangle$.
For the spin-valley half-metal,
$\langle s_v\rangle = 1/2$
for both Fermi pockets. As a result, we readily obtain
\begin{eqnarray}
\label{SVcondSDW}
 \sigma^v=\sigma_0^vG(x),
\end{eqnarray}
where
$\sigma_0^v=\sigma_0^s=\sigma_0/2e$.

\subsection{CDW half-metal}
\label{CDW_current}

In the case of the CDW half-metal, electron-like and hole-like charge
carriers have the same spin projections, while the spin-valley projections
have opposite signs. It is easily proven that in the CDW case, the charge
conductivity is the same as for the SDW phase,
Eq.~\eqref{Electric_SDW},
while the spin and spin-valley conductivities must be interchanged, as
compared to the spin-valley half-metal phase, that is,
\begin{eqnarray}
\label{CDW_cond}
 \sigma_\textrm{E}\!=\!\sigma_0G(x),\quad \sigma^s\!=\! \sigma_0^sG(x),\quad  \sigma^v \!\approx\! 0.
\end{eqnarray}
We can see from
Eqs.~(\ref{CDW_cond})
and~(\ref{SVcondSDW})
that the electric current in our systems carries, besides charge, an
additional quantum number: either spin, or spin-valley projection.

\section{Superconductivity}
\label{sec::SC}

In the half-metal phases under study, we have itinerant electrons in two
Fermi pockets. Therefore, an attractive interaction between these
quasiparticles can give rise to unconventional superconductivity. We
briefly analyze such a possibility. For simplicity, we consider below only
commensurate SDW or CDW ordering.

Let us assume that the effective Hamiltonian of the system can be written as
\begin{equation}
\label{ham}
\hat{H}_{\textrm{eff}}=\hat{H}_{\textrm{HM}}+\hat{H}_{\textrm{\textrm{BCS}}},
\end{equation}
where the first term in the right-hand side,
$\hat{H}_{\textrm{HM}}$,
corresponds either to the spin-valley SDW phase,
Eq.~\eqref{H_SDW},
or to the CDW half-metal,
Eq.~\eqref{H_CDW}.
The second term is a usual BCS attraction. We consider first the CDW phase.
In this case, all electrons in both Fermi-pockets have the same spin
projection
$\sigma=\uparrow$.
Thus, the BCS term can be expressed as
\begin{equation}
\label{CDWsingle_ham}
\hat{H}_{\textrm{BCS}}=-\!\!\sum_{\mathbf{kk}'\alpha\beta}\!
{V^{\alpha\beta}_{\mathbf{kk'}}C^\dag_{\mathbf{k}\uparrow\alpha}
C^\dag_{\mathbf{-k}\uparrow\alpha}
C_{\mathbf{-k'}\uparrow\beta}C_{\mathbf{k'}\uparrow\beta}}\,,
\end{equation}
where
$C^\dag_{\mathbf{k}\uparrow\alpha}$
($C_{\mathbf{k}\uparrow\alpha}$)
are the creation (annihilation) operators of a quasiparticle with momentum
$\mathbf{k}$
and spin projection $\uparrow$ at the Fermi surface pocket
$\alpha=e,h$;
while
$V^{\alpha\beta}_{\mathbf{kk'}}=V^{\beta\alpha}_{\mathbf{kk'}}$
are the corresponding matrix elements of the electron-electron attraction.

The superconducting order parameter is commonly defined as
\begin{equation}
\label{Sc_definition}
\Delta_{\rm sc}^\alpha(\mathbf{k})
=
\sum_{\mathbf{k}' \beta }
{V^{\alpha\beta}_{\mathbf{kk}'}}
	\left\langle
		C^\dag_{\mathbf{k'}\uparrow\beta}
		C^\dag_{\mathbf{-k'}\uparrow\beta}
	\right\rangle.
\end{equation}
In particular, this means that
\begin{equation}
\label{Dsym}
\Delta_{\rm sc}^\alpha(\mathbf{k})
=
-\Delta_{\rm sc}^\alpha(-\mathbf{k})\,.
\end{equation}
Following the standard Bogolyubov approach for the case of two-band
superconductivity~\cite{PhysRevLett.3.552},
we obtain a system of equations for calculating the two superconducting
gaps
\begin{equation}
\label{BCS_P2}
\Delta_{\rm sc}^\alpha(\mathbf{k})=-\sum_{\mathbf{k}'\beta}
{\frac{V^{\alpha\beta}_{\mathbf{kk}'}\Delta_{\rm sc}^\beta(\mathbf{k'})}
{2E^\beta_\mathbf{k'}}}
\tanh{\!\!\left(\frac{E^\beta_\mathbf{k'}}{2T}\right)},
\end{equation}
where
\begin{equation}
\label{Ek}
E^\alpha_\mathbf{k}
=
\sqrt{
	\left(E^{(2)}_{\mathbf{k}\uparrow}-\mu\right)^2
	+
	\Delta_{\rm sc}^\alpha(\mathbf{k})^2}\,.
\end{equation}
In this expression
$E^{(2)}_{\mathbf{k}\uparrow}$
is determined by
Eq.~\eqref{spectrum},
in which the SDW order parameter
$\Delta_\sigma$
should be replaced by the CDW order parameter
$\tilde{\Delta}_\sigma$.
Note that in the case of the CDW half-metal, both gaps
$\Delta_{\rm sc}^{e,h}(\mathbf{k})$
correspond to superconductivity with a spin-polarized supercurrent.

In the case of a usual half-metal, an unconventional superconducting
ordering exists if the matrix element of the electron-electron attraction
obeys certain symmetry
rules~\cite{PhysRevLett.77.3185,PhysRevB.57.557}.
In contrast to a usual half-metal, we have a two-component superconducting
order parameter, one component per one valley. However, the symmetry
analysis of
$V^{\alpha\beta}_{\mathbf{kk}'}$
is very similar to the case of a single-component unconventional
superconductivity. We simply have to demand that the symmetry of the matrix
element should be consistent with the symmetry of the order parameters,
Eq.~\eqref{Dsym}.

For simplicity, let us assume that
$V^{ee}_{\mathbf{kk}'}=V^{hh}_{\mathbf{kk}'}
=
V^{eh}_{\mathbf{kk}'}=V_{\mathbf{kk}'}$.
These assumptions are reasonable, since the difference in the Fermi momenta
of different Fermi pockets is small. From the definition
\begin{eqnarray}
V_{\mathbf{kk}'}
=
\langle \mathbf{k}'\uparrow,-\mathbf{k}'\uparrow|
V|\mathbf{k}\uparrow,-\mathbf{k}\uparrow\rangle.
\end{eqnarray}
Thus, the matrix element must obey the following symmetry
rules~\cite{RevModPhys.63.239,PhysRevB.57.557}
\begin{equation}
\label{sym2}
V_{\mathbf{kk}'}=-V_{\mathbf{-kk}'}=-V_{\mathbf{k-k}'}=V_{\mathbf{-k-k}'}.
\end{equation}
We conclude that the interaction matrix element should have a definite
$\mathbf{k}$-space
dependence, otherwise
$V_{\mathbf{kk}'}=0$
according to
Eq.~\eqref{sym2}
and then superconductivity would be impossible. This non-trivial
$\mathbf{k}$-space
dependent interaction must ensure a correct sign of the sum in the
right-hand side of
Eq.~\eqref{BCS_P2}.
For example, if the matrix element has the form
\begin{eqnarray}
V_{\mathbf{kk}'}=-\frac{\mathbf{kk}'}{k_F^2}V_0,
\end{eqnarray}
then
$V_0$
must be positive and
\begin{eqnarray}
\Delta_{\rm sc}^\alpha(\mathbf{k})
=
f(\mathbf{k})\bar{\Delta}_{\rm sc}^\alpha.
\end{eqnarray}
Here
$\bar{\Delta}_{\rm sc}^\alpha$
is
$\mathbf{k}$-independent
and derived as the usual BCS superconducting gap in the case of the
two-band
model~\cite{PhysRevLett.3.552}.
It depends on the Fermi momenta
$k_F^\alpha$
defined by
Eq.~\eqref{pockets}
and on the interaction parameter
$V_0$.

The above discussion can be easily adopted to the case of the spin-valley
half-metal phase. The only difference is that the spin polarizations of the
two valleys are antiparallel to each other. Consequently, the supercurrent
carries a spin-valley polarization. As for the spin polarization of the
current, it is small, or even zero.

As in the case of the usual BCS treatment, our consideration of the
superconductivity is valid if the superconducting gap is much smaller than
the characteristic Fermi energy of the half-metal state. That is,
\begin{eqnarray}
|\bar{\Delta}_{\rm sc}^\alpha|\ll \Delta_0(1-x/2x_0-\sqrt{1-x/x_0}),
\end{eqnarray}
which in the case of sufficiently high doping,
$x\sim x_0$,
reduces to the condition
$|\bar{\Delta}_{\rm sc}^\alpha|\ll \Delta_0$.

\section{Discussion}
\label{sec::discussion}

Here we have discussed a weak-coupling mechanism of half-metallicity. Since
it does not require a strong electron-electron coupling, it may be
operational in systems composed of light atoms only. For example, the
proposed half-metallicity could exist in systems without transition metals.
Moreover, in addition to the usual half-metal with spin-polarized electrons
at the Fermi surface, we predicted the possible existence of a new phase,
which we referred to as a spin-valley half-metal. This phase is
characterized by the valley quantum number, and the charge carriers at the
Fermi surface are not only spin-polarized but also valley-polarized.
This unique property may be of interest for applications in spintronics,
and the newly emerging field of spin-valley-tronics.

The presented mechanism for the formation of half-metallicity is quite
general, and may be relevant to any material with nesting-driven density
waves. However, here we consider only a specific type of interaction,
namely, short-range electron-electron repulsion,
Eqs.~\eqref{Hint}--\eqref{eq::ham_ex},
with $g$ and
$g_\bot>0$.
In this case, we observe two instabilities of the electronic state: SDW and
CDW. From the former, the spin-valley half-metal state emerges,
Fig.~\ref{valleys}(b),
while the latter one gives rise to the CDW half-metal state,
Fig.~\ref{valleys}(c).
Note that in real materials, a short-range approximation for the
electron-electron coupling is well justified when the system is in a metallic
(or in our case half-metallic) state. In the SDW or CDW insulating state, the long-range interaction could be of significance. However, the use of a more
sophisticated interaction potential does not affect our main results: the
density-wave instability occurs in the system with nesting under the
condition of weak coupling and the ground state of doped system (when the
electron-electron interaction is a short-range one) is the half-metal.

We assume that both the electron and hole sheets of the Fermi surface are
perfectly nested at zero doping. More realistically, the sheets have
non-identical shapes, causing finite denesting even at zero doping. For
example, one sheet may be spherical, while the other may be
elliptical~\cite{Sboychakov_PRB2013_PS_pnict}.

If the zero-doping denesting is sufficiently weak, the range of doping
where
$\partial^2 F^{\rm ic}_0 (x)/\partial x^2 < 0$
shrinks~\cite{Sboychakov_PRB2013_PS_pnict},
but does not disappear. When the sheet shapes differ significantly, one has
$\partial^2 F^{\rm ic}_0 (x)/\partial x^2 > 0$
for all $x$, and the half-metallic states become impossible.

On the other hand, if the sheets are non-spherical, but the zero-doping
nesting is preserved (at
$x=0$
the sheets are identical), our conclusions endure, and only minor
mathematical modifications to the formalism are required (the density of
states acquires a dependence on the spherical angles).

In addition, we assumed the electron-hole symmetry of the ``bare'' (when
the electron-hole coupling is neglected) bands,
Fig.~\ref{valleys1}.
This approximation simplifies the intermediate formulas considerably;
fortunately, it does not trivialize the main results. Straightforward
modifications to the formalism allows one to study a more general model.

It is interesting to note that the model we investigate in this paper is
well-known and was discussed in many research papers. Yet, despite these
efforts,
Hamiltonian~(\ref{ham_summa})
provides an unexpected many-body phase of electronic liquid. This is
associated with the fact that a doped density-wave system has several
states whose energies are almost identical (``stripes", phase separation,
incommensurate density waves). They compete against each other to become
the ``true" ground state. The multiplicity of the competing phases makes a
theoretical description particularly challenging: it is impossible to prove
that no new states will not be added to the list in the future. Thus, to
realize the proposed mechanism in an actual material, a multidisciplinary
study is necessary. In addition to analytical many-body tools, numerical
{\it ab initio}
calculations of Fermi surfaces and other electronic and lattice properties
are highly desirable. Of course, a guidance from the experiment is
indispensable in such a study.

The most striking feature of the half-metal states considered in this paper
is the possibility to observe spin or spin-valley polarized currents. The
corresponding conductivities are significant if the doping is not small and
is of the order of the characteristic value
\begin{eqnarray}
x_0=N_\textrm{F}\Delta_0\sim \Delta_0n_0/\varepsilon_\textrm{F}.
\end{eqnarray}
In this regime, the results obtained are valid at sufficiently low
temperatures,
$T\ll\Delta_0$,
and in the absence of a strong electron scattering,
$1/\tau\ll\Delta_0$.
The absence of magnetic impurities that spoil the spin polarization is also
necessary. We neglected here several perturbations (disorder, spin-orbit
coupling, Umklapp processes). The stability margins of the half-metallic
phases against these factors, as well as their effects on the polarized
currents, should be checked in further studies.

Since the half-metals posses an ungapped Fermi surface, superconductivity
may coexist with these phases. The allowed type of superconductivity is
$p$-wave, with parallel or antiparallel orientations of spin polarizations
on the electronic and hole sheets. When the polarizations are parallel
(antiparallel), the supercurrent, in addition to the electric charge,
carries also spin (spin-valley) quantum.

The electronic phase separation and formation of inhomogeneous states
of electronic matter is an inherent property of systems with imperfect
nesting~\cite{our_chrom2013,Rakhmanov2017}.
A strong long-range Coulomb repulsion suppresses the formation of
inhomogeneous states. We assumed that this Coulomb interaction guarantees
the homogeneity of the electron liquid and neglected the possibility of
phase separation. However, the problem of phase separation in the system
considered here is of interest and deserves a separate analysis because it
makes the phase diagram of the model richer.

The above calculations demonstrate that, among several mean-field states
discussed above, the incommensurate spin-valley half-metal has the lowest
energy, at least for not too strong doping. However, in realistic
$sp$-electron materials the exchange interaction is
small~\cite{sp_exchange}.
Then, the renormalization of the interaction constant for the CDW ordering
Eq.~\eqref{renorm}
is also small. Therefore, the difference in the free energies between the
SDW and CDW phases cannot be large. The difference in the free energy
between the incommensurate and commensurate states is also small if
coupling is weak, as it follows directly from our calculations. It is
reasonable to assume that, in general, factors neglected in our treatment
(temperature, magnetic field, disorder, electron-lattice coupling, etc.)
may change the ground state. However, in any of the studied half-metal
phases, one can observe either spin or spin-valley currents.

To conclude, we discussed the recently proposed weak-coupling mechanism for
half-metallicity, as well as its most immediate consequences. We calculated
the phase diagram for the studied model and explored the connection between
spin conductivity, spin-valley conductivity, and usual electric
conductivity for different phases of the model. We also pointed out that in
our model the half-metallicity may coexist with superconductivity. The
supercurrent in such a superconducting phase would demonstrate nontrivial
spin or spin-valley polarization. The mechanism discussed in this work may
be of importance for the current search for non-toxic
biologically-compatible materials with nontrivial electronic properties.

\section*{Acknowledgments}

This work is partially supported by the JSPS-Russian Foundation for Basic
Research joint Project No.~17-52-50023, and by the Presidium of the Russian
Academy of Sciences (Program I.7). 
F.N. is supported in part by the MURI Center for Dynamic Magneto-Optics via
the Air Force Office of Scientific Research (AFOSR) (FA9550-14-1-0040),
Army Research Office (ARO) (Grant No.~W911NF-18-1-0358), Asian Office of
Aerospace Research and Development (AOARD) (Grant No.~FA2386-18-1-4045),
Japan Science and Technology Agency (JST) (the ImPACT program and CREST
Grant No.~JPMJCR1676), Japan Society for the Promotion of Science (JSPS)
(JSPS-RFBR Grant No.~17-52-50023, and JSPS-FWO Grant No.~VS.059.18N),
RIKEN-AIST Challenge Research Fund, and the John Templeton Foundation.
K.I.K.  and A.V.R. acknowledge the support of the Russian Foundation for
Basic Research, Projects No.~17-02-00323 and No.~17-02-00135. 
A.V.R. is also grateful to the Skoltech NGP Program (Skoltech-MIT joint
project) for additional support.

\bibliographystyle{apsrevlong_no_issn_url}

\begin{thebibliography}{42}
\expandafter\ifx\csname natexlab\endcsname\relax\def\natexlab#1{#1}\fi
\expandafter\ifx\csname bibnamefont\endcsname\relax
  \def\bibnamefont#1{#1}\fi
\expandafter\ifx\csname bibfnamefont\endcsname\relax
  \def\bibfnamefont#1{#1}\fi
\expandafter\ifx\csname citenamefont\endcsname\relax
  \def\citenamefont#1{#1}\fi

\bibitem[{\citenamefont{de~Groot et~al.}(1983)\citenamefont{de~Groot, Mueller,
  van Engen, and Buschow}}]{first_half_met1983}
\bibinfo{author}{\bibfnamefont{R.~A.} \bibnamefont{de~Groot}},
  \bibinfo{author}{\bibfnamefont{F.~M.} \bibnamefont{Mueller}},
  \bibinfo{author}{\bibfnamefont{P.~G.} \bibnamefont{van Engen}},
  \bibnamefont{and} \bibinfo{author}{\bibfnamefont{K.~H.~J.}
  \bibnamefont{Buschow}}, {``}\bibinfo{title}{New Class of Materials:
  Half-Metallic Ferromagnets},{''} \bibinfo{journal}{Phys. Rev. Lett.}
  \textbf{\bibinfo{volume}{50}}, \bibinfo{pages}{2024} (\bibinfo{year}{1983}).

\bibitem[{\citenamefont{Katsnelson et~al.}(2008)\citenamefont{Katsnelson,
  Irkhin, Chioncel, Lichtenstein, and de~Groot}}]{half_met_review2008}
\bibinfo{author}{\bibfnamefont{M.~I.} \bibnamefont{Katsnelson}},
  \bibinfo{author}{\bibfnamefont{V.~Y.} \bibnamefont{Irkhin}},
  \bibinfo{author}{\bibfnamefont{L.}~\bibnamefont{Chioncel}},
  \bibinfo{author}{\bibfnamefont{A.~I.} \bibnamefont{Lichtenstein}},
  \bibnamefont{and} \bibinfo{author}{\bibfnamefont{R.~A.}
  \bibnamefont{de~Groot}}, {``}\bibinfo{title}{Half-metallic ferromagnets: From
  band structure to many-body effects},{''} \bibinfo{journal}{Rev. Mod. Phys.}
  \textbf{\bibinfo{volume}{80}}, \bibinfo{pages}{315} (\bibinfo{year}{2008}).

\bibitem[{\citenamefont{Hu}(2012)}]{hu2012half}
\bibinfo{author}{\bibfnamefont{X.}~\bibnamefont{Hu}},
  {``}\bibinfo{title}{Half-Metallic Antiferromagnet as a Prospective Material
  for Spintronics},{''} \bibinfo{journal}{Adv. Mater.}
  \textbf{\bibinfo{volume}{24}}, \bibinfo{pages}{294} (\bibinfo{year}{2012}).

\bibitem[{\citenamefont{{\v{Z}}uti{\'{c}}
  et~al.}(2004)\citenamefont{{\v{Z}}uti{\'{c}}, Fabian, and
  Das~Sarma}}]{review_spintronics2004}
\bibinfo{author}{\bibfnamefont{I.}~\bibnamefont{{\v{Z}}uti{\'{c}}}},
  \bibinfo{author}{\bibfnamefont{J.}~\bibnamefont{Fabian}}, \bibnamefont{and}
  \bibinfo{author}{\bibfnamefont{S.}~\bibnamefont{Das~Sarma}},
  {``}\bibinfo{title}{Spintronics: Fundamentals and applications},{''}
  \bibinfo{journal}{Rev. Mod. Phys.} \textbf{\bibinfo{volume}{76}},
  \bibinfo{pages}{323} (\bibinfo{year}{2004}).

\bibitem[{\citenamefont{Hanssen et~al.}(1990)\citenamefont{Hanssen, Mijnarends,
  Rabou, and Buschow}}]{nimnsb_exp1990}
\bibinfo{author}{\bibfnamefont{K.~E. H.~M.} \bibnamefont{Hanssen}},
  \bibinfo{author}{\bibfnamefont{P.~E.} \bibnamefont{Mijnarends}},
  \bibinfo{author}{\bibfnamefont{L.~P. L.~M.} \bibnamefont{Rabou}},
  \bibnamefont{and} \bibinfo{author}{\bibfnamefont{K.~H.~J.}
  \bibnamefont{Buschow}}, {``}\bibinfo{title}{Positron-annihilation study of
  the half-metallic ferromagnet NiMnSb: Experiment},{''}
  \bibinfo{journal}{Phys. Rev. B} \textbf{\bibinfo{volume}{42}},
  \bibinfo{pages}{1533} (\bibinfo{year}{1990}).

\bibitem[{\citenamefont{Park et~al.}(1998)\citenamefont{Park, Vescovo, Kim,
  Kwon, Ramesh, and Venkatesan}}]{lasrmno_half_met_exp1998}
\bibinfo{author}{\bibfnamefont{J.-H.} \bibnamefont{Park}},
  \bibinfo{author}{\bibfnamefont{E.}~\bibnamefont{Vescovo}},
  \bibinfo{author}{\bibfnamefont{H.-J.} \bibnamefont{Kim}},
  \bibinfo{author}{\bibfnamefont{C.}~\bibnamefont{Kwon}},
  \bibinfo{author}{\bibfnamefont{R.}~\bibnamefont{Ramesh}}, \bibnamefont{and}
  \bibinfo{author}{\bibfnamefont{T.}~\bibnamefont{Venkatesan}},
  {``}\bibinfo{title}{Direct evidence for a half-metallic ferromagnet},{''}
  \bibinfo{journal}{Nature} \textbf{\bibinfo{volume}{392}},
  \bibinfo{pages}{794} (\bibinfo{year}{1998}).

\bibitem[{\citenamefont{Ji et~al.}(2001)\citenamefont{Ji, Strijkers, Yang,
  Chien, Byers, Anguelouch, Xiao, and Gupta}}]{cro2_half_met_exp2001}
\bibinfo{author}{\bibfnamefont{Y.}~\bibnamefont{Ji}},
  \bibinfo{author}{\bibfnamefont{G.~J.} \bibnamefont{Strijkers}},
  \bibinfo{author}{\bibfnamefont{F.~Y.} \bibnamefont{Yang}},
  \bibinfo{author}{\bibfnamefont{C.~L.} \bibnamefont{Chien}},
  \bibinfo{author}{\bibfnamefont{J.~M.} \bibnamefont{Byers}},
  \bibinfo{author}{\bibfnamefont{A.}~\bibnamefont{Anguelouch}},
  \bibinfo{author}{\bibfnamefont{G.}~\bibnamefont{Xiao}}, \bibnamefont{and}
  \bibinfo{author}{\bibfnamefont{A.}~\bibnamefont{Gupta}},
  {``}\bibinfo{title}{Determination of the Spin Polarization of Half-Metallic
  ${\mathrm{CrO}}_{2}$ by Point Contact Andreev Reflection},{''}
  \bibinfo{journal}{Phys. Rev. Lett.} \textbf{\bibinfo{volume}{86}},
  \bibinfo{pages}{5585} (\bibinfo{year}{2001}).

\bibitem[{\citenamefont{Jourdan et~al.}(2014)\citenamefont{Jourdan,
  Min{\'{a}}¡r, Braun, Kronenberg, Chadov, Balke, Gloskovskii, Kolbe, Elmers,
  Sch{\"{o}}nhense et~al.}}]{co2mnsi_half_met_exp2014}
\bibinfo{author}{\bibfnamefont{M.}~\bibnamefont{Jourdan}},
  \bibinfo{author}{\bibfnamefont{J.}~\bibnamefont{Min{\'{a}}¡r}},
  \bibinfo{author}{\bibfnamefont{J.}~\bibnamefont{Braun}},
  \bibinfo{author}{\bibfnamefont{A.}~\bibnamefont{Kronenberg}},
  \bibinfo{author}{\bibfnamefont{S.}~\bibnamefont{Chadov}},
  \bibinfo{author}{\bibfnamefont{B.}~\bibnamefont{Balke}},
  \bibinfo{author}{\bibfnamefont{A.}~\bibnamefont{Gloskovskii}},
  \bibinfo{author}{\bibfnamefont{M.}~\bibnamefont{Kolbe}},
  \bibinfo{author}{\bibfnamefont{H.}~\bibnamefont{Elmers}},
  \bibinfo{author}{\bibfnamefont{G.}~\bibnamefont{Sch{\"{o}}nhense}},
  \bibnamefont{et~al.}, {``}\bibinfo{title}{Direct observation of
  half-metallicity in the Heusler compound Co{$_2$}MnSi},{''}
  \bibinfo{journal}{Nat. Commun.} \textbf{\bibinfo{volume}{5}},
  \bibinfo{pages}{3974} (\bibinfo{year}{2014}).

\bibitem{M. P. Ghimire} M. P. Ghimire, L.-H. Wu, and Xiao Hu, "Possible half-metallic antiferromagnetism in an iridium double-perovskite material", Phys. Rev. B {\bf 93}, 134421 (2016).

\bibitem[{\citenamefont{Du et~al.}(2012)\citenamefont{Du, Sanvito, and
  Smith}}]{metal_free_hm2012}
\bibinfo{author}{\bibfnamefont{A.}~\bibnamefont{Du}},
  \bibinfo{author}{\bibfnamefont{S.}~\bibnamefont{Sanvito}}, \bibnamefont{and}
  \bibinfo{author}{\bibfnamefont{S.~C.} \bibnamefont{Smith}},
  {``}\bibinfo{title}{First-Principles Prediction of Metal-Free Magnetism and
  Intrinsic Half-Metallicity in Graphitic Carbon Nitride},{''}
  \bibinfo{journal}{Phys. Rev. Lett.} \textbf{\bibinfo{volume}{108}},
  \bibinfo{pages}{197207} (\bibinfo{year}{2012}).

\bibitem[{\citenamefont{Hashmi and Hong}(2014)}]{meta_free_hm2014}
\bibinfo{author}{\bibfnamefont{A.}~\bibnamefont{Hashmi}} \bibnamefont{and}
  \bibinfo{author}{\bibfnamefont{J.}~\bibnamefont{Hong}},
  {``}\bibinfo{title}{Metal free half metallicity in 2D system: structural and
  magnetic properties of g-C$_4$N$_3$ on BN},{''} \bibinfo{journal}{Sci. Rep.}
  \textbf{\bibinfo{volume}{4}}, \bibinfo{pages}{4374} (\bibinfo{year}{2014}).

\bibitem[{\citenamefont{Son et~al.}(2006)\citenamefont{Son, Cohen, and
  Louie}}]{son2006half}
\bibinfo{author}{\bibfnamefont{Y.-W.} \bibnamefont{Son}},
  \bibinfo{author}{\bibfnamefont{M.~L.} \bibnamefont{Cohen}}, \bibnamefont{and}
  \bibinfo{author}{\bibfnamefont{S.~G.} \bibnamefont{Louie}},
  {``}\bibinfo{title}{Half-metallic graphene nanoribbons},{''}
  \bibinfo{journal}{Nature} \textbf{\bibinfo{volume}{444}},
  \bibinfo{pages}{347} (\bibinfo{year}{2006}).

\bibitem[{\citenamefont{Kan et~al.}(2012)\citenamefont{Kan, Hu, Xiao, Lu, Deng,
  Yang, and Su}}]{kan2012half}
\bibinfo{author}{\bibfnamefont{E.}~\bibnamefont{Kan}},
  \bibinfo{author}{\bibfnamefont{W.}~\bibnamefont{Hu}},
  \bibinfo{author}{\bibfnamefont{C.}~\bibnamefont{Xiao}},
  \bibinfo{author}{\bibfnamefont{R.}~\bibnamefont{Lu}},
  \bibinfo{author}{\bibfnamefont{K.}~\bibnamefont{Deng}},
  \bibinfo{author}{\bibfnamefont{J.}~\bibnamefont{Yang}}, \bibnamefont{and}
  \bibinfo{author}{\bibfnamefont{H.}~\bibnamefont{Su}},
  {``}\bibinfo{title}{Half-metallicity in organic single porous sheets},{''}
  \bibinfo{journal}{Journal of the American Chemical Society}
  \textbf{\bibinfo{volume}{134}}, \bibinfo{pages}{5718} (\bibinfo{year}{2012}).

\bibitem[{\citenamefont{Huang et~al.}(2010)\citenamefont{Huang, Si, Lee, Zhao,
  Wu, Gu, and Duan}}]{huang2010intrinsic}
\bibinfo{author}{\bibfnamefont{B.}~\bibnamefont{Huang}},
  \bibinfo{author}{\bibfnamefont{C.}~\bibnamefont{Si}},
  \bibinfo{author}{\bibfnamefont{H.}~\bibnamefont{Lee}},
  \bibinfo{author}{\bibfnamefont{L.}~\bibnamefont{Zhao}},
  \bibinfo{author}{\bibfnamefont{J.}~\bibnamefont{Wu}},
  \bibinfo{author}{\bibfnamefont{B.-L.} \bibnamefont{Gu}}, \bibnamefont{and}
  \bibinfo{author}{\bibfnamefont{W.}~\bibnamefont{Duan}},
  {``}\bibinfo{title}{Intrinsic half-metallic BN--C nanotubes},{''}
  \bibinfo{journal}{Applied Physics Letters} \textbf{\bibinfo{volume}{97}},
  \bibinfo{pages}{043115} (\bibinfo{year}{2010}).

\bibitem[{\citenamefont{Soriano and Fern\'andez-Rossier}(2010)}]{soriano2010}
\bibinfo{author}{\bibfnamefont{D.}~\bibnamefont{Soriano}} \bibnamefont{and}
  \bibinfo{author}{\bibfnamefont{J.}~\bibnamefont{Fern\'andez-Rossier}},
  {``}\bibinfo{title}{Spontaneous persistent currents in a quantum spin Hall
  insulator},{''} \bibinfo{journal}{Phys. Rev. B}
  \textbf{\bibinfo{volume}{82}}, \bibinfo{pages}{161302}
  (\bibinfo{year}{2010}).

\bibitem[{\citenamefont{Klauk}(2010)}]{plastic_electr2010}
\bibinfo{author}{\bibfnamefont{H.}~\bibnamefont{Klauk}},
  {``}\bibinfo{title}{Organic thin-film transistors},{''}
  \bibinfo{journal}{Chem. Soc. Rev.} \textbf{\bibinfo{volume}{39}},
  \bibinfo{pages}{2643} (\bibinfo{year}{2010}).

\bibitem[{\citenamefont{Avouris et~al.}(2007)\citenamefont{Avouris, Chen, and
  Perebeinos}}]{Avouris2007}
\bibinfo{author}{\bibfnamefont{P.}~\bibnamefont{Avouris}},
  \bibinfo{author}{\bibfnamefont{Z.}~\bibnamefont{Chen}}, \bibnamefont{and}
  \bibinfo{author}{\bibfnamefont{V.}~\bibnamefont{Perebeinos}},
  {``}\bibinfo{title}{Carbon-based electronics},{''} \bibinfo{journal}{Nat.
  Nanotechnol.} \textbf{\bibinfo{volume}{2}}, \bibinfo{pages}{605}
  (\bibinfo{year}{2007}).

\bibitem[{\citenamefont{Rozhkov et~al.}(2011)\citenamefont{Rozhkov, Giavaras,
  Bliokh, Freilikher, and Nori}}]{meso_review}
\bibinfo{author}{\bibfnamefont{A.}~\bibnamefont{Rozhkov}},
  \bibinfo{author}{\bibfnamefont{G.}~\bibnamefont{Giavaras}},
  \bibinfo{author}{\bibfnamefont{Y.~P.} \bibnamefont{Bliokh}},
  \bibinfo{author}{\bibfnamefont{V.}~\bibnamefont{Freilikher}},
  \bibnamefont{and} \bibinfo{author}{\bibfnamefont{F.}~\bibnamefont{Nori}},
  {``}\bibinfo{title}{Electronic properties of mesoscopic graphene structures:
  Charge confinement and control of spin and charge transport},{''}
  \bibinfo{journal}{Phys. Rep.} \textbf{\bibinfo{volume}{503}},
  \bibinfo{pages}{77 } (\bibinfo{year}{2011}).

\bibitem[{\citenamefont{Sa-Ke et~al.}(2014)\citenamefont{Sa-Ke, Hong-Yu,
  Yong-Hong, and Jun}}]{chinese_phys_silicene2014}
\bibinfo{author}{\bibfnamefont{W.}~\bibnamefont{Sa-Ke}},
  \bibinfo{author}{\bibfnamefont{T.}~\bibnamefont{Hong-Yu}},
  \bibinfo{author}{\bibfnamefont{Y.}~\bibnamefont{Yong-Hong}},
  \bibnamefont{and} \bibinfo{author}{\bibfnamefont{W.}~\bibnamefont{Jun}},
  {``}\bibinfo{title}{Spin and valley half metal induced by staggered potential
  and magnetization in silicene},{''} \bibinfo{journal}{Chin. Phys. B}
  \textbf{\bibinfo{volume}{23}}, \bibinfo{pages}{017203}
  (\bibinfo{year}{2014}).

\bibitem[{\citenamefont{Rozhkov et~al.}(2016)\citenamefont{Rozhkov, Sboychakov,
  Rakhmanov, and Nori}}]{bilayer_review2016}
\bibinfo{author}{\bibfnamefont{A.}~\bibnamefont{Rozhkov}},
  \bibinfo{author}{\bibfnamefont{A.}~\bibnamefont{Sboychakov}},
  \bibinfo{author}{\bibfnamefont{A.}~\bibnamefont{Rakhmanov}},
  \bibnamefont{and} \bibinfo{author}{\bibfnamefont{F.}~\bibnamefont{Nori}},
  {``}\bibinfo{title}{Electronic properties of graphene-based bilayer
  systems},{''} \bibinfo{journal}{Phys. Rep.} \textbf{\bibinfo{volume}{648}},
  \bibinfo{pages}{1 } (\bibinfo{year}{2016}).

\bibitem[{\citenamefont{Rozhkov et~al.}(2017)\citenamefont{Rozhkov, Rakhmanov,
  Sboychakov, Kugel, and Nori}}]{PhysRevLett.119.107601}
\bibinfo{author}{\bibfnamefont{A.~V.} \bibnamefont{Rozhkov}},
  \bibinfo{author}{\bibfnamefont{A.~L.} \bibnamefont{Rakhmanov}},
  \bibinfo{author}{\bibfnamefont{A.~O.} \bibnamefont{Sboychakov}},
  \bibinfo{author}{\bibfnamefont{K.~I.} \bibnamefont{Kugel}}, \bibnamefont{and}
  \bibinfo{author}{\bibfnamefont{F.}~\bibnamefont{Nori}},
  {``}\bibinfo{title}{Spin-Valley Half-Metal as a Prospective Material for Spin
  Valleytronics},{''} \bibinfo{journal}{Phys. Rev. Lett.}
  \textbf{\bibinfo{volume}{119}}, \bibinfo{pages}{107601}
  (\bibinfo{year}{2017}).

\bibitem[{\citenamefont{Rice}(1970)}]{Rice}
\bibinfo{author}{\bibfnamefont{T.~M.} \bibnamefont{Rice}},
  {``}\bibinfo{title}{Band-Structure Effects in Itinerant
  Antiferromagnetism},{''} \bibinfo{journal}{Phys. Rev. B}
  \textbf{\bibinfo{volume}{2}}, \bibinfo{pages}{3619} (\bibinfo{year}{1970}).

\bibitem[{\citenamefont{Rakhmanov et~al.}(2013)\citenamefont{Rakhmanov,
  Rozhkov, Sboychakov, and Nori}}]{our_chrom2013}
\bibinfo{author}{\bibfnamefont{A.~L.} \bibnamefont{Rakhmanov}},
  \bibinfo{author}{\bibfnamefont{A.~V.} \bibnamefont{Rozhkov}},
  \bibinfo{author}{\bibfnamefont{A.~O.} \bibnamefont{Sboychakov}},
  \bibnamefont{and} \bibinfo{author}{\bibfnamefont{F.}~\bibnamefont{Nori}},
  {``}\bibinfo{title}{Phase separation of antiferromagnetic ground states in
  systems with imperfect nesting},{''} \bibinfo{journal}{Phys. Rev. B}
  \textbf{\bibinfo{volume}{87}}, \bibinfo{pages}{075128}
  (\bibinfo{year}{2013}).

\bibitem[{\citenamefont{Gorbatsevich et~al.}(1992)\citenamefont{Gorbatsevich,
  Kopaev, and Tokatly}}]{tokatly1992}
\bibinfo{author}{\bibfnamefont{A.}~\bibnamefont{Gorbatsevich}},
  \bibinfo{author}{\bibfnamefont{Y.}~\bibnamefont{Kopaev}}, \bibnamefont{and}
  \bibinfo{author}{\bibfnamefont{I.}~\bibnamefont{Tokatly}},
  {``}\bibinfo{title}{Band theory of phase stratification},{''}
  \bibinfo{journal}{Zh. Eksp. Teor. Fiz.} \textbf{\bibinfo{volume}{101}},
  \bibinfo{pages}{971} (\bibinfo{year}{1992}), \bibinfo{note}{[Sov. Phys. JETP
  {\bf 74}, 521 (1992)]}.

\bibitem[{\citenamefont{Eremin and Chubukov}(2010)}]{eremin_chub2010}
\bibinfo{author}{\bibfnamefont{I.}~\bibnamefont{Eremin}} \bibnamefont{and}
  \bibinfo{author}{\bibfnamefont{A.~V.} \bibnamefont{Chubukov}},
  {``}\bibinfo{title}{Magnetic degeneracy and hidden metallicity of the
  spin-density-wave state in ferropnictides},{''} \bibinfo{journal}{Phys. Rev.
  B} \textbf{\bibinfo{volume}{81}}, \bibinfo{pages}{024511}
  (\bibinfo{year}{2010}).

\bibitem[{\citenamefont{Rakhmanov et~al.}(2012)\citenamefont{Rakhmanov,
  Rozhkov, Sboychakov, and Nori}}]{PrlOur}
\bibinfo{author}{\bibfnamefont{A.~L.} \bibnamefont{Rakhmanov}},
  \bibinfo{author}{\bibfnamefont{A.~V.} \bibnamefont{Rozhkov}},
  \bibinfo{author}{\bibfnamefont{A.~O.} \bibnamefont{Sboychakov}},
  \bibnamefont{and} \bibinfo{author}{\bibfnamefont{F.}~\bibnamefont{Nori}},
  {``}\bibinfo{title}{Instabilities of the $AA$-Stacked Graphene Bilayer},{''}
  \bibinfo{journal}{Phys. Rev. Lett.} \textbf{\bibinfo{volume}{109}},
  \bibinfo{pages}{206801} (\bibinfo{year}{2012}).

\bibitem[{\citenamefont{Sboychakov
  et~al.}(2013{\natexlab{a}})\citenamefont{Sboychakov, Rozhkov, Rakhmanov, and
  Nori}}]{PrbOur}
\bibinfo{author}{\bibfnamefont{A.~O.} \bibnamefont{Sboychakov}},
  \bibinfo{author}{\bibfnamefont{A.~V.} \bibnamefont{Rozhkov}},
  \bibinfo{author}{\bibfnamefont{A.~L.} \bibnamefont{Rakhmanov}},
  \bibnamefont{and} \bibinfo{author}{\bibfnamefont{F.}~\bibnamefont{Nori}},
  {``}\bibinfo{title}{Antiferromagnetic states and phase separation in doped
  $AA$-stacked graphene bilayers},{''} \bibinfo{journal}{Phys. Rev. B}
  \textbf{\bibinfo{volume}{88}}, \bibinfo{pages}{045409}
  (\bibinfo{year}{2013}{\natexlab{a}}).

\bibitem[{\citenamefont{Sboychakov
  et~al.}(2013{\natexlab{b}})\citenamefont{Sboychakov, Rakhmanov, Rozhkov, and
  Nori}}]{PrbROur}
\bibinfo{author}{\bibfnamefont{A.~O.} \bibnamefont{Sboychakov}},
  \bibinfo{author}{\bibfnamefont{A.~L.} \bibnamefont{Rakhmanov}},
  \bibinfo{author}{\bibfnamefont{A.~V.} \bibnamefont{Rozhkov}},
  \bibnamefont{and} \bibinfo{author}{\bibfnamefont{F.}~\bibnamefont{Nori}},
  {``}\bibinfo{title}{Metal-insulator transition and phase separation in doped
  $AA$-stacked graphene bilayer},{''} \bibinfo{journal}{Phys. Rev. B}
  \textbf{\bibinfo{volume}{87}}, \bibinfo{pages}{121401}
  (\bibinfo{year}{2013}{\natexlab{b}}).

\bibitem[{\citenamefont{Sboychakov
  et~al.}(2013{\natexlab{c}})\citenamefont{Sboychakov, Rozhkov, Kugel,
  Rakhmanov, and Nori}}]{Sboychakov_PRB2013_PS_pnict}
\bibinfo{author}{\bibfnamefont{A.~O.} \bibnamefont{Sboychakov}},
  \bibinfo{author}{\bibfnamefont{A.~V.} \bibnamefont{Rozhkov}},
  \bibinfo{author}{\bibfnamefont{K.~I.} \bibnamefont{Kugel}},
  \bibinfo{author}{\bibfnamefont{A.~L.} \bibnamefont{Rakhmanov}},
  \bibnamefont{and} \bibinfo{author}{\bibfnamefont{F.}~\bibnamefont{Nori}},
  {``}\bibinfo{title}{Electronic phase separation in iron pnictides},{''}
  \bibinfo{journal}{Phys. Rev. B} \textbf{\bibinfo{volume}{88}},
  \bibinfo{pages}{195142} (\bibinfo{year}{2013}{\natexlab{c}}).

\bibitem[{\citenamefont{Akzyanov et~al.}(2014)\citenamefont{Akzyanov,
  Sboychakov, Rozhkov, Rakhmanov, and Nori}}]{PrbVOur}
\bibinfo{author}{\bibfnamefont{R.~S.} \bibnamefont{Akzyanov}},
  \bibinfo{author}{\bibfnamefont{A.~O.} \bibnamefont{Sboychakov}},
  \bibinfo{author}{\bibfnamefont{A.~V.} \bibnamefont{Rozhkov}},
  \bibinfo{author}{\bibfnamefont{A.~L.} \bibnamefont{Rakhmanov}},
  \bibnamefont{and} \bibinfo{author}{\bibfnamefont{F.}~\bibnamefont{Nori}},
  {``}\bibinfo{title}{$AA$-stacked bilayer graphene in an applied electric
  field: Tunable antiferromagnetism and coexisting exciton order
  parameter},{''} \bibinfo{journal}{Phys. Rev. B}
  \textbf{\bibinfo{volume}{90}}, \bibinfo{pages}{155415}
  (\bibinfo{year}{2014}).

\bibitem[{\citenamefont{Sboychakov et~al.}(2017)\citenamefont{Sboychakov,
  Rakhmanov, Kugel, Rozhkov, and Nori}}]{prb_sl2017}
\bibinfo{author}{\bibfnamefont{A.~O.} \bibnamefont{Sboychakov}},
  \bibinfo{author}{\bibfnamefont{A.~L.} \bibnamefont{Rakhmanov}},
  \bibinfo{author}{\bibfnamefont{K.~I.} \bibnamefont{Kugel}},
  \bibinfo{author}{\bibfnamefont{A.~V.} \bibnamefont{Rozhkov}},
  \bibnamefont{and} \bibinfo{author}{\bibfnamefont{F.}~\bibnamefont{Nori}},
  {``}\bibinfo{title}{Magnetic field effects in electron systems with imperfect
  nesting},{''} \bibinfo{journal}{Phys. Rev. B} \textbf{\bibinfo{volume}{95}},
  \bibinfo{pages}{014203} (\bibinfo{year}{2017}).

\bibitem[{\citenamefont{Moreo et~al.}(1999)\citenamefont{Moreo, Yunoki, and
  Dagotto}}]{moreo1999}
\bibinfo{author}{\bibfnamefont{A.}~\bibnamefont{Moreo}},
  \bibinfo{author}{\bibfnamefont{S.}~\bibnamefont{Yunoki}}, \bibnamefont{and}
  \bibinfo{author}{\bibfnamefont{E.}~\bibnamefont{Dagotto}},
  {``}\bibinfo{title}{Phase Separation Scenario for Manganese Oxides and
  Related Materials},{''} \bibinfo{journal}{Science}
  \textbf{\bibinfo{volume}{283}}, \bibinfo{pages}{2034} (\bibinfo{year}{1999}).

\bibitem[{\citenamefont{Dagotto}(2003)}]{dagotto_book}
\bibinfo{author}{\bibfnamefont{E.}~\bibnamefont{Dagotto}},
  \emph{\bibinfo{title}{Nanoscale Phase Separation and Colossal
  Magnetoresistance}}, Springer Series in Solid-State Sciences
  (\bibinfo{publisher}{Springer}, \bibinfo{address}{Berlin},
  \bibinfo{year}{2003}).

\bibitem[{\citenamefont{Dagotto et~al.}(2003)\citenamefont{Dagotto, Burgy, and
  Moreo}}]{dagotto_phasep2003}
\bibinfo{author}{\bibfnamefont{E.}~\bibnamefont{Dagotto}},
  \bibinfo{author}{\bibfnamefont{J.}~\bibnamefont{Burgy}}, \bibnamefont{and}
  \bibinfo{author}{\bibfnamefont{A.}~\bibnamefont{Moreo}},
  {``}\bibinfo{title}{Nanoscale phase separation in colossal magnetoresistance
  materials: lessons for the cuprates?},{''} \bibinfo{journal}{Solid State
  Commun.} \textbf{\bibinfo{volume}{126}}, \bibinfo{pages}{9 }
  (\bibinfo{year}{2003}), \bibinfo{note}{proceedings of the High-Tc
  Superconductivity Workshop}.

\bibitem[{\citenamefont{Igoshev et~al.}(2010)\citenamefont{Igoshev, Timirgazin,
  Katanin, Arzhnikov, and Irkhin}}]{IrkhinPRB2010}
\bibinfo{author}{\bibfnamefont{P.~A.} \bibnamefont{Igoshev}},
  \bibinfo{author}{\bibfnamefont{M.~A.} \bibnamefont{Timirgazin}},
  \bibinfo{author}{\bibfnamefont{A.~A.} \bibnamefont{Katanin}},
  \bibinfo{author}{\bibfnamefont{A.~K.} \bibnamefont{Arzhnikov}},
  \bibnamefont{and} \bibinfo{author}{\bibfnamefont{V.~Y.}
  \bibnamefont{Irkhin}}, {``}\bibinfo{title}{Incommensurate magnetic order and
  phase separation in the two-dimensional Hubbard model with nearest- and
  next-nearest-neighbor hopping},{''} \bibinfo{journal}{Phys. Rev. B}
  \textbf{\bibinfo{volume}{81}}, \bibinfo{pages}{094407}
  (\bibinfo{year}{2010}).

\bibitem[{\citenamefont{Lorenzana et~al.}(2001)\citenamefont{Lorenzana,
  Castellani, and Di~Castro}}]{di_castro1}
\bibinfo{author}{\bibfnamefont{J.}~\bibnamefont{Lorenzana}},
  \bibinfo{author}{\bibfnamefont{C.}~\bibnamefont{Castellani}},
  \bibnamefont{and}
  \bibinfo{author}{\bibfnamefont{C.}~\bibnamefont{Di~Castro}},
  {``}\bibinfo{title}{Phase separation frustrated by the long-range Coulomb
  interaction. I. Theory},{''} \bibinfo{journal}{Phys. Rev. B}
  \textbf{\bibinfo{volume}{64}}, \bibinfo{pages}{235127}
  (\bibinfo{year}{2001}).

\bibitem[{\citenamefont{Bianconi et~al.}(2015)\citenamefont{Bianconi, Poccia,
  Sboychakov, Rakhmanov, and Kugel}}]{bianconi2015intrinsic}
\bibinfo{author}{\bibfnamefont{A.}~\bibnamefont{Bianconi}},
  \bibinfo{author}{\bibfnamefont{N.}~\bibnamefont{Poccia}},
  \bibinfo{author}{\bibfnamefont{A.}~\bibnamefont{Sboychakov}},
  \bibinfo{author}{\bibfnamefont{A.}~\bibnamefont{Rakhmanov}},
  \bibnamefont{and} \bibinfo{author}{\bibfnamefont{K.}~\bibnamefont{Kugel}},
  {``}\bibinfo{title}{Intrinsic arrested nanoscale phase separation near a
  topological Lifshitz transition in strongly correlated two-band metals},{''}
  \bibinfo{journal}{Supercond. Sci. Technol.} \textbf{\bibinfo{volume}{28}},
  \bibinfo{pages}{024005} (\bibinfo{year}{2015}).

\bibitem[{\citenamefont{Han}(2013)}]{han2013modern}
\bibinfo{author}{\bibfnamefont{F.}~\bibnamefont{Han}}, \emph{\bibinfo{title}{A
  Modern Course in the Quantum Theory of Solids}} (\bibinfo{publisher}{World
  Scientific}, \bibinfo{year}{2013}), ISBN \bibinfo{isbn}{9789814417143}.

\bibitem[{\citenamefont{Suhl et~al.}(1959)\citenamefont{Suhl, Matthias, and
  Walker}}]{PhysRevLett.3.552}
\bibinfo{author}{\bibfnamefont{H.}~\bibnamefont{Suhl}},
  \bibinfo{author}{\bibfnamefont{B.~T.} \bibnamefont{Matthias}},
  \bibnamefont{and} \bibinfo{author}{\bibfnamefont{L.~R.}
  \bibnamefont{Walker}}, {``}\bibinfo{title}{Bardeen-Cooper-Schrieffer Theory
  of Superconductivity in the Case of Overlapping Bands},{''}
  \bibinfo{journal}{Phys. Rev. Lett.} \textbf{\bibinfo{volume}{3}},
  \bibinfo{pages}{552} (\bibinfo{year}{1959}).

\bibitem[{\citenamefont{Pickett}(1996)}]{PhysRevLett.77.3185}
\bibinfo{author}{\bibfnamefont{W.~E.} \bibnamefont{Pickett}},
  {``}\bibinfo{title}{Single Spin Superconductivity},{''}
  \bibinfo{journal}{Phys. Rev. Lett.} \textbf{\bibinfo{volume}{77}},
  \bibinfo{pages}{3185} (\bibinfo{year}{1996}).

\bibitem[{\citenamefont{Rudd and Pickett}(1998)}]{PhysRevB.57.557}
\bibinfo{author}{\bibfnamefont{R.~E.} \bibnamefont{Rudd}} \bibnamefont{and}
  \bibinfo{author}{\bibfnamefont{W.~E.} \bibnamefont{Pickett}},
  {``}\bibinfo{title}{Single-spin superconductivity: Formulation and
  Ginzburg-Landau theory},{''} \bibinfo{journal}{Phys. Rev. B}
  \textbf{\bibinfo{volume}{57}}, \bibinfo{pages}{557} (\bibinfo{year}{1998}).

\bibitem[{\citenamefont{Sigrist and Ueda}(1991)}]{RevModPhys.63.239}
\bibinfo{author}{\bibfnamefont{M.}~\bibnamefont{Sigrist}} \bibnamefont{and}
  \bibinfo{author}{\bibfnamefont{K.}~\bibnamefont{Ueda}},
  {``}\bibinfo{title}{Phenomenological theory of unconventional
  superconductivity},{''} \bibinfo{journal}{Rev. Mod. Phys.}
  \textbf{\bibinfo{volume}{63}}, \bibinfo{pages}{239} (\bibinfo{year}{1991}).

\bibitem[{\citenamefont{Rakhmanov et~al.}(2017)\citenamefont{Rakhmanov, Kugel,
  Kagan, Rozhkov, and Sboychakov}}]{Rakhmanov2017}
\bibinfo{author}{\bibfnamefont{A.~L.} \bibnamefont{Rakhmanov}},
  \bibinfo{author}{\bibfnamefont{K.~I.} \bibnamefont{Kugel}},
  \bibinfo{author}{\bibfnamefont{M.~Y.} \bibnamefont{Kagan}},
  \bibinfo{author}{\bibfnamefont{A.~V.} \bibnamefont{Rozhkov}},
  \bibnamefont{and} \bibinfo{author}{\bibfnamefont{A.~O.}
  \bibnamefont{Sboychakov}}, {``}\bibinfo{title}{Inhomogeneous electron states
  in the systems with imperfect nesting},{''} \bibinfo{journal}{JETP Letters}
  \textbf{\bibinfo{volume}{105}}, \bibinfo{pages}{806} (\bibinfo{year}{2017}).

\bibitem{sp_exchange} E. \c{S}a\c{s}oglu, I. Galanakis, C. Friedrich, S. Bl\"{u}gel, "\textit{Ab initio} calculation of the effective on-site Coulomb interaction parameters for half-metallic magnets," Phys. Rev. B {\bf 88}, 134402 (2013).


\end{thebibliography}

\end{document}